# Spectroscopic origin of giant anomalous Hall effect in an interwoven magnetic kagome metal

Erjian Cheng[1,#,*], Kaipu Wang[2,#], Yiqing Hao[3,#], Wenqing Chen[4,#], Hengxin Tan[5,#], Zongkai Li[2,#], Meixiao Wang[2], Wenli Gao[6], Di Wu[6], Shuaishuai Sun[7], Tianping Ying[7], Simin Nie[8], Yiwei Li[9], Walter Schnelle[1], Houke Chen[10], Xingjiang Zhou[6,7], Ralf Koban[1], Yulin Chen[2,10], Binghai Yan[5], Yi-feng Yang[7,6,11,*], Weida Wu[4,*], Zhongkai Liu[2,*] and Claudia Felser[1,*]

[1]*Max Planck Institute for Chemical Physics of Solids, 01187 Dresden, Germany*

[2]*School of Physical Science and Technology, ShanghaiTech Laboratory for Topological Physics, ShanghaiTech University, 201210 Shanghai, China*

[3]*Neutron Scattering Division, Oak Ridge National Laboratory, P.0. Box 2008, Oak Ridge, Tennessee, 37831, USA*

[4]*Department of Physics & Astronomy, Rutgers University, Piscataway, NJ 08854, USA*

[5]*Department of Condensed Matter Physics, Weizmann Institute of Science, Rehovot 7610001, Israel*

[6]*University of Chinese Academy of Sciences, Chinese Academy of Sciences, 100190 Beijing, China*

[7]*Beijing National Laboratory for Condensed Matter Physics, Institute of Physics, Chinese Academy of Sciences, Beijing 100190, China*

[8]*Department of Mechanical Engineering, Stanford University, 94305 Stanford, California, USA*

[9]*Institute for Advanced Studies (IAS), Wuhan University, 430072 Wuhan, China*

[10]*Department of Physics, University of Oxford, Oxford OX1 3PU, UK*

[11]*Songshan Lake Materials Laboratory, Dongguan, Guangdong 523808, China*

## Abstract

The discovery of a giant anomalous Hall effect (AHE) and its novel mechanism holds significant promise for advancing both fundamental research and practical applications. Magnetic kagome lattice materials are uniquely suited for studying the AHE due to their interplay between electronic structure, topology, and magnetism. However, the geometric frustration inherent in kagome lattices often limits the configuration and tunability of magnetic order. Here, we present a new design strategy for kagome-lattice materials with emergent magnetism, exemplified by the magnetic kagome metal $TbTi_3Bi_4$, which features interwoven magnetic Tb zigzag chains and non-magnetic Ti kagome bilayers. This material exhibits a record-high anomalous Hall conductivity (AHC) of $10^5$ $\Omega^{-1}cm^{-1}$. Spectroscopy measurements reveal a large band folding gap observed via angle-resolved photoemission spectroscopy, coexisting spin-density-wave (SDW) order detected through spin-polarized scanning tunneling spectroscopy, and a spiral magnetic order with large magnetic moments identified by neutron diffraction. These findings highlight a strong electron-magnetic coupling between itinerant charges and ordered magnetic moments, offering a spectroscopic explanation for the giant AHC in $TbTi_3Bi_4$. This work establishes a pathway for innovative material design strategies, unlocking new possibilities for future exploration and applications in quantum and spintronic technologies.

## Introduction

The anomalous Hall effect (AHE), an intriguing phenomenon in condensed matter physics, has garnered significant interest from both theoretical and experimental perspectives [1]. Unlike the ordinary Hall effect, which arises from the Lorentz force bending charge carrier trajectories perpendicular to an applied magnetic field, the AHE originates from two distinct microscopic mechanisms: extrinsic processes caused by scattering effects and an intrinsic mechanism related to the Berry curvature [1]. In the intrinsic case, the resulting Hall conductivity set by the Berry curvature is expected to be less than $e^2/ha$ in three dimensions (where $h$ is Planck's constant and $a$ is a typical lattice constant), corresponding to values on the order of $\sigma_{xy}$ = $10^2$–$10^3$ $\Omega^{-1}$ $cm^{-1}$. In contrast, the contribution from extrinsic mechanisms has no such upper threshold, for example, the skew scattering effect—asymmetric electron scattering due to the spin-orbital coupling (SOC) at impurities [1]. Furthermore, the spin-chirality skew-scattering mechanism has been proposed as a potential driver for the giant AHC [2–4]. Specifically, the giant AHC is attributed to magnetic textures associated with a finite scalar spin chirality, $\chi_{ijk} = \boldsymbol{S}_i \cdot (\boldsymbol{S}_j \times \boldsymbol{S}_k)$, where $\boldsymbol{S}_{i,j,k}$ represent the spins. This insight offers a pathway to achieving a giant AHC, even though the underlying mechanism remains poorly understood. Thus, investigating its origin would not only deepen the theoretical understanding but also open avenues for material design for potential applications.

The kagome lattice, made up of corner-sharing triangles could simultaneously host topologically non-trivial electronic states, electron correlations and magnetism and has proved to be a fertile platform for studying intriguing quantum phenomena [5–7]. Moreover, the introduction of magnetic ordering in kagome-lattice materials may bring nontrivial topological phases, such as kagome Chern magnets and time-reversal symmetry breaking topological Weyl semimetals [8–10]; further, the magnetic ordering may help establish the emergent charge ordering in kagome-lattice materials [11,12]. At this moment the studies are limited on materials with magnetic atoms form kagome lattices, which limits the configuration of the magnetic ground state (due to frustration) and individual tuning capability of the magnetic moment, thus preventing us from the exploration of new physics and design and optimization of magneto-transport performance.

Recently, the titanium-based bilayer kagome metals $LnTi_3Bi_4$ ($Ln$ = rare-earth elements) family has emerged as such interesting candidates, which shows various and tunable magnetism by integrating rare-earth magnetic layers with titanium-based kagome layers [13–19]. In this study, we implemented such novel approach by synthesizing $TbTi_3Bi_4$, which features interwoven magnetic Tb zigzag chains and non-magnetic kagome Ti layers. The quasi-one-dimensional (1D) magnetic Tb spin chains, combined with the kagome-lattice electronic structure, synergistically produce novel physics not observed in other kagome compounds, including various instabilities (Supplementary Fig. 1) and magneto-transport behavior. When a magnetic field is applied along the Tb zigzag chains, a giant AHC on the order of $10^5\ \Omega^{-1}\ cm^{-1}$ has been observed. Our spectroscopy measurements provide spectroscopy origin of such AHC: 1. Using angle-resolved photoemission spectroscopy (ARPES), we uncover a quasi-1D Tb band folding with a wavevector of (1/3, 0, 0), accompanied by a substantial hybridization gap of approximately 90 meV. 2. Our spin-polarized scanning tunneling microscopy (SP-STM) reveals a stripe order with the same wavevector, suggesting the formation of a SDW order for the first time in kagome metals. 3. Our neutron diffraction measurements resolve a unique magnetic structure and discover large ordering moments. Integrating these results suggest that the large band folding gap is driven by a strong electron-magnetic coupling between itinerant charges and the large ordered moments rather than Fermi surface nesting and therefore signals a spectroscopic origin for the record-high AHC in $TbTi_3Bi_4$.

## Results

### Interweaved structure, magnetic properties and giant AHC in $TbTi_3Bi_4$

$TbTi_3Bi_4$ crystallizes in an orthorhombic structure within the *Fmmm* (No. 69) space group. This structure comprises alternating layers of $Ti_3Bi_4$, $LnBi_2$, and Bi along the *c* axis, as shown in Fig. 1a. Unlike the $D_{6h}$ symmetry observed in kagome metals $AM_3Sb_5$ (where $A$ = K, Rb, Cs, $M$ = V, Ti), the *Ln* atoms form quasi-1D zigzag chains running along the *a* axis (Fig. 1b) [20–22]. This leads to an orthorhombic structure in the $LnBi_2$ layer, resulting in a reduced crystalline symmetry ($D_{2h}$). Quasi-1D spin chains can naturally possess magnetic frustration and strong quantum fluctuations [23,24], which, intriguingly, lead to complex magnetic textures and transport behaviors in $TbTi_3Bi_4$ featuring an interwoven kagome lattice. Figure 1c shows the magnetization with magnetic field applied along three perpendicular crystal axes. With decreasing temperature, magnetization displays a hump at low temperature, implying strong crystal-field effect with two-level splitting as resolved by specific heat measurements (Supplementary Fig. 2a). A sudden drop at 20.4 K can be clearly identified for all axis, corresponding to a first magnetic ordering at $T_{N1}$. For magnetic field applied along the *a* axis, a second magnetic ordering can be clearly resolved at $T_{N2}$ = 3 K. Curie-Weiss type fits to the paramagnetic susceptibility demonstrate antiferromagnetic (AFM) interactions for magnetic field parallel to the *b* and *c* axis, and ferromagnetic interaction for the *a* axis, suggestive of strong magnetic anisotropy of $TbTi_3Bi_4$ (Supplementary Fig. 2b). Figure 1d displays the profiles of magnetoresistance (MR) with different measured configurations at 2 K. When magnetic field is applied along the *b* and *c* axis, MR shows a large and positive value of ~ 150% at 9 T. However, when applying magnetic field only parallel to the *a* axis, the MR shows a negative value of ~ -50%, featuring multiple anomalies, which are also prominent in Hall resistivity, corresponding to metamagnetic transitions [Fig. 1e]. It is noteworthy that the negative MR observed persists up to 300 K, much higher than the long-range ordering temperature $T_{N1}$ (Supplementary Fig. 4), indicating that it is the easy damping of the spin fluctuation components perpendicular to the long-range ordering vector (*viz*, the *a* axis) when applying a field along this direction.

More intriguingly, when a magnetic field exceeding 3 T (the saturation field) is applied along the chain direction, the Hall resistivity exhibits a linear behavior (middle panel of Fig. 1e), in stark contrast to other measured configurations (Supplementary Fig. 3). To separate the contributions, the empirical relation $\rho_{yx} = \rho_{yx}^{O} + \rho_{yx}^{A} = R_0\mu_0 H + \rho_{yx}^{A}$ has been employed, where $\rho_{yx}^{O} = R_0\mu_0 H$ ($R_0$ is a constant) and $\rho_{yx}^{A}$ represent the ordinary and anomalous Hall contributions, respectively. The resulting anomalous Hall resistivity reveals three distinct regimes, labeled I (0.4 T < $\mu_0 H$ < 0.7 T during field down-sweeping), II (1 T < $\mu_0 H$ < 2.4 T), and III ($\mu_0 H$ > 3 T), corresponding to the low-field plateau, the 1/3 magnetization plateau, and the saturated magnetization, respectively, as schematized in the magnetization phase diagram (Fig. 1f). The spikes observed at the metamagnetic transitions are likely attributed to domain wall effects, similar to those reported in $GdTi_3Bi_4$ [19]. The AHC has been also calculated, as displayed in the lower panel in Fig. 1e. At low-field plateau (0.4 T < $\mu_0 H$ < 0.7 T during field down-sweeping), the longitudinal conductivity ($\sigma_{xx}$) and AHC at 2 K are 4.4 × $10^5$ and 6.7 × $10^4$ $\Omega^{-1}$ $cm^{-1}$, respectively, with an anomalous Hall angle (AHA) of 15.2%. At 1/3 plateau (1 T < $\mu_0 H$ < 2.4 T), the $\sigma_{xx}$ and AHC are 3.3 × $10^5$ and 5.1 × $10^4$ $\Omega^{-1}$ $cm^{-1}$ with an AHA of 15.1%. At regime III, the $\sigma_{xx}$ and AHC are 4.8 × $10^5$ and 1.5 × $10^5$ $\Omega^{-1}$ $cm^{-1}$ with an AHA of 31.1%. The AHC in Phase I is particularly noteworthy, as it corresponds to a small moment of approximately 0.4 $\mu_B$ of $Tb^{3+}$ in magnetization measurements. Neutron scattering experiments (discussed later) confirm that $TbTi_3Bi_4$ is an antiferromagnet. Its elliptical spiral magnetic structure and low-field AHC characteristics may suggest the potential altermagnetic nature of $TbTi_3Bi_4$ in low fields [25]. Figure 1g shows the AHC versus $\sigma_{xx}$ for $TbTi_3Bi_4$ with other reported various materials for comparison. The longitudinal conductivity clearly falls within the empirical intrinsic regime, where the Berry curvature mechanism predominates. Notably, the observed AHC significantly exceeds the intrinsic AHE threshold derived from momentum-space considerations. It far surpasses the estimated quantization limit ($e^2/ha$ ~ 659 $\Omega^{-1}$ $cm^{-1}$ with $a$ = 5.8666 Å) for $TbTi_3Bi_4$ in three dimensions. These findings indicate the exotic and intricate coupling between magnetic texture and electrons, which motivate us to explore the electronic band and magnetic structures.

**Electronic band folding and large quasi-1D hybridization gap in $TbTi_3Bi_4$**

To scrutinize the electronic band structure, we performed ARPES experiments. The overall electronic structure of $TbTi_3Bi_4$ measured by ARPES is detailed in Supplementary Note 5, including the characteristic kagome electronic structure from Ti 3*d* orbitals [Dirac cones at Fermi level ($E_F$) and multiple van Hove singularities (vHSs) close to $E_F$ (Supplementary Figs. 6,7,10)]. Here we focus the transition of the electronic structure across $T_{N1}$. Figure 2a shows the Fermi surface (FS) mapping measured at 6.3 K ($T_{N2} < T < T_{N1}$) with the projected BZ (see the definition of the full BZ in Supplementary Fig. 6a), from which three types of FSs can be resolved: the triangular FSs near zone corners ($\overline{K}$, $\overline{K'}$) originate from the Ti 3*d* orbitals (see Supplementary Note 6, same below); the quasi-1D FS, which shows minimal dispersion along the $k_x$ direction, comes from mostly Tb 5*d* (especially $d_{xz}$) orbitals, as Tb forms a quasi-1D chain; The inner ellipse comes from mostly Bi 6*p* orbitals. These observations are consistent with the results of band structure calculations (Supplementary Note 6) and previous measurements of other $Ln Ti_3Bi_4$ systems [13–19].

Upon closely examining the FS originating mostly from the Tb 5*d* orbitals, we observed significant signature of band folding from 30 ($T > T_{N2}$) to 6.3 K (Fig. 2b). The shadow-like FS indicated by the green arrows in Fig. 2b appears/weakens at 6.3/30 K and has the shape of the FS with a nesting vector $\boldsymbol{q}$ = (1/3, 0, 0). In addition, the quasi-1D FS connected by the $\boldsymbol{q}$ vector (red arrows in Fig. 2b) loses/regains spectral weight at 6.3/30 K, suggesting the gap opening behavior due to band folding. The band folding is best visualized in the comparison of dispersion across the Tb bands measured at 6.3 [Fig. 2c (ii)] and 30 K [Fig. 2c (iii)], where clear evidence of band folding with a folding vector $\boldsymbol{q}$ labeled, and a band hybridization gap [labelled as $\Delta_{gap}$ in Fig. 2c(ii)] opening at $k_x = \pm$ 0.185 Å$^{-1}$ ($\pm 1/2\boldsymbol{q}$) could be observed. Notably, band hybridization occurs not only in the Tb 5*d* bands but also in the Bi 6*p* and Ti 3*d* bands, as indicated by the signature of band folding and hybridization gap shown in the cuts measured using linearly vertical photon polarization [Fig. 2d (ii), also see Supplementary Note 7 for more data on the folding band/hybridization gap]. The same dispersion measured at 30 K (above $T_N$) shows the disappearance of the band folding and the hybridization gap [Fig. 2d (iii)].

Our systematic temperature-dependent measurement of the cut in Fig. 2c reveals the closing of the hybridization gap across $T_{N1} \sim 20$ K (Figs. 2f and g), suggesting a close relationship between the band folding and the magnetic phase transition. Furthermore, the folding-induced hybridization band gap is also quasi-1D due to the quasi-1D electronic structure from the Tb 5*d* bands. Figure 2e shows the measured dispersion along the nested FS, capturing the hybridization gap at all $k_y$ values (see Supplementary Figs. 15–17 for more data). From the summarized hybridization gap versus the $k_y$ plot (Fig. 2h), we find the hybridization gap is largest (~90 meV) for $k_y <$ -0.25 Å$^{-1}$ and becomes smaller for -0.25 Å$^{-1} < k_y <$ -0.15 Å$^{-1}$. The existence of hybridization gap explains the suppression of intensity of quasi-1D FS at $E_F$ at large $k_y$ ($|k_y| > 0.25$ Å$^{-1}$, see Fig. 2a). Moreover, at all $k_z$ values, the band folding and the hybridization gap remain similar (see Supplementary Fig. 21). The gap exists in whole $k_z$ space indicates that the folding vector only exists in the $k_x$–$k_y$ planes. More ARPES data and analysis on the hybridization gap could be found in Supplementary Notes 7,8. Supplementary Note 9 presents lower-temperature data at 1.7 K below $T_{N2}$, revealing an emergent new band that likely results from band folding from another magnetic ordering ***q'*** or band shifting.

**Spin-density-wave (SDW) order and magnetic structure of $TbTi_3Bi_4$**

The band folding and large hybridization gap is accompanied with a spin density wave order as observed by our spin-polarized (SP) STM measurements. Figure 3 displays the non-spin polarized (NSP) and SP-STM measurements (more data and analysis could be found in Supplementary Note 10). Figure 3a shows the NSP-STM topograph of $TbTi_3Bi_4$ with Tb termination measured at 4.2 K along with its corresponding Fourier transforms (FT), clearly revealing Bragg peaks ($Q_{Bragg}$). Additional ordering ($Q_{rec}$) was observed in the local density of states (LDOS) maps from NSP-STM measured at $E_F$-40 meV and their corresponding FT results at 4.2 K but disappears at 24 K [Figs. 3b(i,ii), the schematic of reconstruction was shown in Figs. 3c and b(iii)]. These peaks may be related to a charge ordering due to surface reconstructions but the ***q*** vector is not the one found from ARPES. Meanwhile the cryogenic transmission electron microscope measurements did not provide any bulk reconstruction, thus a bulk charge density wave order could be ruled out (Supplementary Note 11). On the other hand, the SP-STM experiments confirm the presence of the spin density wave. The LDOS map and FT at 4.55 K [Fig. 3d(i)] reveals the Bragg peaks and the similar reconstruction peaks observed in NSP-STM. In addition, a stripe like patterns were observed in SP-STM and corresponding peaks found in the FT results. These stripes correspond to a triple lattice reconstruction along the Tb chain [Fig. 3c, and d(iii)] and consistent with the SDW ordering vector $\boldsymbol{q}$ = (1/3, 0, 0) found in the ARPES measurements. At high temperature in the paramagnetic state, the peaks of SDW disappear [Fig. 3d(ii)], demonstrating that SDW ordering only occurs below $T_{N1}$. The SDW was observed for the first time in all the kagome latticed materials. On the other hand, the band folding gap of about 90 meV observed in $TbTi_3Bi_4$ is unusually large for an SDW-driven mechanism, for which the expected mean field gap size $\Delta = 3.53kT_N$ [35] is only about 6 meV for $T_N = 20.4$ K, which challenges the conventional SDW order as the sole origin of the gap. In addition, the existence of the hybridization gap in the bands away from $E_F$ (Fig. 2e) also suggests other key factors additional to the FS nesting for the band folding.

**Spiral magnetic structure and large magnetic moment of $TbTi_3Bi_4$**

To further explore the magnetic order of $TbTi_3Bi_4$, we performed temperature-dependent neutron diffraction measurement, as displayed in Fig. 4. Neutron diffraction measurements performed at 1.5 K reveals strong magnetic peaks at $\boldsymbol{q_1}$ = (±0.36, ±0.29, $L$), with $L$ = 0, 2, 4, … (Figs. 4a,b), corresponding to the incommensurate magnetic propagation vector $\boldsymbol{k_1}$ = (0.36, 0.29, 0). This result is partially consistent with the observed $\boldsymbol{q} = (1/3, 0, 0)$ from ARPES and SP-STM, showing a triple lattice periodicity along the Tb chain direction ($a$ axis). To determine the magnetic structure, symmetry analysis was performed using $\boldsymbol{k_1} = (0.36, 0.29, 0)$ and the parent space group *Fmmm* (No. 69). Two irreducible representations were found, but only $\Gamma_1$ successfully describes the observed magnetic peak intensities. The magnetic structure is best explained by a twinned single-$\boldsymbol{q}$ ellipsoid spiral, with the easy axis along the $a$ axis. The ordered magnetic moment projections along the $a$, $b$, and $c$ axis are $\boldsymbol{m}_a = 10.1(1)$ $\mu_B$/Tb, $\boldsymbol{m}_b = 3.5(5)$ $\mu_B$/Tb and $\boldsymbol{m}_c = 1.9(3)$ $\mu_B$/Tb, respectively. In addition to the primary magnetic order at $\boldsymbol{k_1}$, weaker magnetic peaks were observed at $\boldsymbol{q_2} = (\pm0.08, \pm0.18, L)$, with $L = 1, 3, 5, \ldots$ (Fig. 4c), corresponding to a secondary magnetic propagation vector $\boldsymbol{k_2} = (\pm0.08, \pm0.18, 1)$. The magnetic structure of $\boldsymbol{k_2}$ is best described by a cosine-wave modulated structure, with its easy axis also along the $a$ axis. The ordered magnetic moment size is $\boldsymbol{m}_a = 1.9(1)$ $\mu_B$/Tb. Temperature-dependent measurements were taken for both $\boldsymbol{k_1}$ and $\boldsymbol{k_2}$ propagation vectors, as shown in Fig. 4d, at $\boldsymbol{q_1} = (0.36, 0.29, 6)$ and $\boldsymbol{q_2} = (0.08, 0.18, 3)$, respectively. The $\boldsymbol{q_1}$ peak shows a single-phase transition at $T_N = 20.4$ K, consistent with susceptibility measurements as well as the evolution of the hybridization gap size from ARPES measurements (Fig. 4d). In contrast, the $\boldsymbol{q_2}$ peak exhibits multiple phase transitions at 8, 13 and 16 K. The distinct temperature dependencies of $\boldsymbol{k_1}$ and $\boldsymbol{k_2}$ suggest that the two magnetic wave vectors originate from different sources. The variation between neutron diffraction and SP-STM may be due to the difference between bulk and surface, or the inaccurate $\boldsymbol{q}$ determined from SP-STM measurements, or the quasi-1D nature of the Tb conduction bands. The $\boldsymbol{q}_2$ involves out-of-plane magnetic ordering, which may lead to the $\bar{\Gamma}$-$\bar{Z}$ folding and explains the band structure change across $T_{N2}$ from ARPES measurements (Supplementary Note 9).

## Discussions

Summarizing all above measurements suggest unusual magnetism in $TbTi_3Ti_4$ with coexisting elliptical spiral magnetic structure of large moments and itinerant SDW order of quasi-1D conduction bands. The observation of a SDW has not been reported in kagome-lattice materials previously, making $TbTi_3Ti_4$ a highly unusual case. The SDW has not been reported in other $R$$Ti_3Bi_4$ siblings (for example, $R$ = Gd, Nb, Sm, Eu), even though they have the same vHSs from Ti as well as quasi-1D Fermi surface from rare earth element, suggesting the SDW is not solely driven by the nesting of the Fermi surface, but strongly influenced by local magnetic moment.

The large ordered moments and their coupling with conduction electrons suggest a Kondo lattice scenario that may be described by the following typical Hamiltonian:

$$H = \sum_{\mathbf{k}s} \epsilon_{\mathbf{k}} c^\dagger_{\mathbf{k}s} c_{\mathbf{k}s} + J_K \sum_i \boldsymbol{S}_i \cdot \boldsymbol{s}_i = \sum_{\mathbf{k}s} \epsilon_{\mathbf{k}} c^\dagger_{\mathbf{k}s} c_{\mathbf{k}s} + J_K \sum_{\mathbf{q}} \boldsymbol{S}_{\mathbf{q}} \cdot \boldsymbol{s}_{-\mathbf{q}}, \quad (1)$$

where $\boldsymbol{S}_i$ ($\boldsymbol{S}_q$) is the spin operator of the local moments in real (momentum) space and $\boldsymbol{s}_{\mathbf{q}} = \sum_{\mathbf{k}} c^\dagger_{\mathbf{k}+\mathbf{q}s} \frac{\tau_{ss'}}{2} c_{\mathbf{k}s'}$ is the spin density operator of the conduction electrons. The Kondo coupling $J_K$ naturally induces an effective RKKY interaction between local moments that may explain their magnetic ordering. The large band folding gap may also be understood by replacing $\boldsymbol{S}_q$ by the mean-

field order parameter $M_{\pm\mathbf{Q}}/g\mu_B$ for $\mathbf{q} = \pm\mathbf{Q}$ ($M_{-\mathbf{Q}} = M_{\mathbf{Q}}^*$). This gives the mean-field Hamiltonian in the magnetically ordered state:

$$H = \sum_{\mathbf{k}s}\left(\epsilon_{\mathbf{k}}c_{\mathbf{k}s}^{\dagger}c_{\mathbf{k}s} + \frac{J_K M_{\mathbf{Q}}}{2g\mu_B}sc_{\mathbf{k}s}^{\dagger}c_{\mathbf{k+Q},s} + h.c.\right), \tag{2}$$

where $\mathbf{Q}$ is the ordering wave vector and the spins $s = \pm$ are defined according to the direction of the ordered moments. For the Tb chain, the conduction electrons are scattered between $\mathbf{k}$ and $\mathbf{k} + \mathbf{Q}$ along the chain direction, causing the observed hybridization gap of Tb bands. In matrix form, we have

$$H = \sum_{\mathbf{k}s}\psi_{\mathbf{k}s}^{\dagger}\begin{pmatrix}\epsilon_{\mathbf{k}} & sh \\ sh^* & \epsilon_{\mathbf{k+Q}}\end{pmatrix}\psi_{\mathbf{k}s}, \tag{3}$$

where $\psi_{\mathbf{k}s} = \left(c_{\mathbf{k}s}^{\dagger}, c_{\mathbf{k+Q},s}^{\dagger}\right)$, $h = \frac{J_K M_{\mathbf{Q}}}{2g\mu_B}$, and the sum is over the folded Brillouin zone. Diagonalizing the above matrix gives the hybridized dispersions:

$$E_{\mathbf{k}s}^{\pm} = \frac{\epsilon_{\mathbf{k}} + \epsilon_{\mathbf{k+Q}}}{2} \pm \sqrt{\left(\frac{\epsilon_{\mathbf{k}} - \epsilon_{\mathbf{k+Q}}}{2}\right)^2 + |h|^2}, \tag{4}$$

which predicts a hybridization gap $\Delta = 2|h| = \frac{J_K|M_{\mathbf{Q}}|}{g\mu_B}$ at $\epsilon_{\mathbf{k}} = \epsilon_{\mathbf{k+Q}}$. Thus, the hybridization gap is proportional to the Kondo interaction and the magnitude of the ordered moment. In the case of $TbTi_3Bi_4$, the large hybridization gap reflects a strong electron-magnetic coupling between the conduction bands and the large ordered moments, which provides a microscopic mechanism for the large AHC. Our measurements reveal the spectroscopic origin of giant anomalous Hall effect in $TbTi_3Bi_4$.

Finally, although we have not detected any related structural reconstruction in the bulk of $TbTi_3Bi_4$, our thermal expansion measurement found a discontinuity in the thermal expansion coefficient along the Tb chain direction (Supplementary Fig. 29a), similar to the discontinuity observed from the charge-density-wave transition in the $CsV_3Sb_5$ case (Supplementary Fig. 29c). Such observation further suggests the coupling between magnetism and lattice degree of freedom across the magnetic transition may also play a role. The magnetostriction experiments further confirms this coupling (Supplementary Fig. 29d).

## Conclusions

In summary, we systematically investigated the titanium-based bilayer kagome metal $TbTi_3Bi_4$, uncovering a record-high AHC. Through various spectroscopy measurements, we revealed an unusual magnetic order with coexisting SDW and large ordering moments, and a substantially large band hybridization gap than SDW expectation. These suggest a Kondo lattice scenario, where the gap is mainly induced by strong electron-magnetic coupling between conduction electrons and large ordered moments, which play synergistically in boosting magneto-transport properties and lead to the giant AHC in $TbTi_3Bi_4$. The fact that no comparable AHC has been reported in other $R$Ti$_3$Bi$_4$ compounds and quantum magnets further supports this scenario. Our discovery not only uncovers a key mechanism for achieving giant AHE in kagome-lattice materials but also underscores the transformative potential of this breakthrough for next-generation electronic and quantum technologies.

## Methods

### Sample preparation

$TbTi_3Bi_4$ single crystals were grown using a self-flux method with an elemental ratio of Tb:Ti:Bi of 1.2:3:20. Tb (99.95%), Ti (99.99%), and Bi (99.999%) were cut into small pieces, mixed, and placed in an alumina crucible. The crucible was then sealed in a quartz tube under partial argon pressure. The sealed tube was heated to 800 °C over 12 hours and maintained at that temperature for 24 hours. It was then slowly cooled to 400 °C at a rate of 2 °C per hour. Single crystals were obtained by removing the flux through centrifugation.

### ARPES measurements

High-resolution ARPES measurements were performed at beamlines BL5-2 of the Stanford Synchrotron Radiation Light Source (SSRL) and beamline BL03U of Shanghai Synchrotron Radiation Facility (SSRF, proposal number: S-XV-ST-6370A) with Scienta DA30 analyzer. The photon-energy ranges of data acquisition are 40–80 eV. The samples were cleaved *in situ* at ~7 K and measured in ultrahigh vacuum with a base pressure of better than $5 \times 10^{-11}$ Torr. The energy and momentum resolution were 10 meV and 0.1°, respectively. Ultra-low temperature, high-resolution laser-based ARPES measurements were performed on home-built setups (hν = 6.994 eV) at Institute of Physics, Chinese Academy of Sciences. The samples were cleaved *in situ* at ~2 K and measured under ultrahigh vacuum below $5 \times 10^{-11}$ Torr. Data were collected by a DA30L analyzer. The total convolved energy and angle resolutions were 2 meV and 0.1°, respectively.

### STM measurements

The scanning tunneling microscopy (STM) measurements have been performed using a Unisoku USM-1300 system, with a base pressure of $1.0 \times 10^{-10}$ Torr. The samples were cleaved mechanically *in situ*, and then immediately inserted into the STM head. Topographic images were obtained with Pt/Ir tips with $V = 1$ V, $I = 200$ pA. $dI/dV$ spectra were collected using a standard lock-in technique at a frequency of 973.137 Hz. For spin-polarized scanning tunneling microscopy (SP-STM) experiments, the spin-polarized tip is functionalized by repeatedly scanning the nonmagnetic tungsten tip on cleaved surfaces of FeTe single crystals. The spin-polarized tip is confirmed by visualizing the in-plane collinear antiferromagnetic order on FeTe.

### DFT calculation

Our calculations are performed using the projector augmented wave method [36,37] as implemented in the Vienna *ab initio* Simulation package [38,39]. Experimental lattice structure is adopted in our calculations. The exchange-correlation functional is treated within a generalized gradient approximation parametrized by Perdew, Burke, and Ernzerhof [40]. In the self-consistent calculations, the cutoff energy for the plane-wave expansion is 500 eV, and the *k*-point sampling grid of the Brillouin zone is 5×6×7. To simulate the paramagnetic state, the 4*f* electrons of Tb are treated used as core electrons.

**Electrical, thermodynamic, and dilatometry measurements**

For electrical transport measurements, a single crystal was cut into a bar shape. A standard six-probe method was used for the longitudinal resistivity and transverse Hall measurements. Electrical transport data were collected in a physical property measurement system (PPMS, Quantum Design). Magnetic susceptibility and specific heat measurements were performed in a magnetic property measurement system (MPMS, Quantum Design) and a PPMS, respectively. Thermal expansion measurements were conducted in a PPMS using a home-built capacitive dilatometer as discussed in Ref. [41].

**Neutron diffraction**

Neutron diffraction experiment was conducted at HB-3A DEMAND at High Flux Isotope Reactor at Oak Ridge National Lab. One piece of $TbTi_3Bi_4$ single crystal of dimension 5 × 3 × 0.2 $mm^3$ was aligned in (0*KL*) scatting plane and cooled down to 1.5 K. The wavelength used was 1.541 Å. Data were reduced using MANTID and ReTIA. Symmetry analysis used SARAh, magnetic structure refinement used Fullprof.

## Data availability

The data that support the findings of this study are available from the corresponding authors upon request.

## References

[1] N. Nagaosa, J. Sinova, S. Onoda, A. H. MacDonald, N. P. Ong, Anomalous Hall effect. *Rev. Mod. Phys.* **82**, 1539−1592 (2010).

[2] Yang, S.-Y. *et al*. Giant, unconventional anomalous Hall effect in the metallic frustrated magnet candidate, $KV_3Sb_5$. *Sci. Adv.* **6**, eabb6003 (2020).

[3] Fujishiro, Y. *et al*. Giant anomalous Hall effect from spin-chirality scattering in a chiral magnet. *Nat. Commun.* **12**, 317 (2021).

[4] Ishizuka, H. & Nagaosa, N. Large anomalous Hall effect and spin Hall effect by spin-cluster scattering in the strong-coupling limit. *Phys. Rev. B* **103**, 235148 (2021).

[5] Yin, J.-X., Lian, B. & Hasan, M. Z. Topological kagome magnets and superconductors. *Nature*

**612**, 647–657 (2022).

[6] Wang, Y. J., Wu, H., Mccandless, G. T., Chan, J. Y. & Ali, M. N. Quantum states and intertwining phases in kagome materials. *Nat. Rev. Phys.* **5**, 635–658 (2023).

[7] Zhou, Y., Kanoda, K., & Ng, T.-K. Quantum spin liquid states. *Rev. Mod. Phys*. **89**, 025003 (2017).

[8] Yin, J.-X. *et al*. Quantum-limit Chern topological magnetism in $TbMn_6Sn_6$. *Nature* **583**, 533–536 (2020).

[9] Yin, J.-X., Lian, B. & Hasan, M. Z. Topological kagome magnets and superconductors. *Nature* **612**, 647–657 (2022).

[10] Liu, D. F. *et al*. Magnetic Weyl semimetal phase in a Kagomé crystal. *Science* **365**, 1282–1285 (2019).

[11] Teng, X. *et al*. Discovery of charge density wave in a kagome lattice antiferromagnet. *Nature* **609**, 490–495 (2022).

[12] Teng, X. *et al*., Magnetism and charge density wave order in kagome FeGe. *Nat. Phys.* **19**, 814–822 (2023).

[13] Ortiz, B. R. *et al.* Evolution of highly anisotropic magnetism in the titanium-based kagome metals $LnTi_3Bi_4$ (Ln: La···$Gd^{3+}$, $Eu^{2+}$, $Yb^{2+}$). *Chem. Mater.* **35**, 9756–9773 (2023).

[14] Chen, L. *et al.* Tunable magnetism in titanium-based kagome metals by rare-earth engineering and high pressure. *Commun. Mater.* **5**, 73 (2024).

[15] Zheng, Z. *et al.* Anisotropic magnetism and band evolution induced by ferromagnetic phase transition in titanium-based kagome ferromagnet $SmTi_3Bi_4$. *Sci. China Phys. Mech. Astron.* **67**, 267411 (2024).

[16] Sakhya, A. P. *et al.* Observation of multiple flat bands and topological Dirac states in a new titanium based slightly distorted kagome metal $YbTi_3Bi_4$. Preprint at https://arxiv.org/abs/2309.01176 (2023).

[17] Jiang, Z. *et al.* Direct observation of topological surface states in the layered kagome lattice with broken time-reversal symmetry. Preprint at https://arxiv.org/abs/2309.01579 (2023).

[18] Hu, Y. *et al.* Magnetic-coupled electronic landscape in bilayer-distorted titanium-based kagome metals. *Phys. Rev. B* **110**, L121114 (2024).

[19] Cheng, E. J. *et al*. Striped magnetization plateau and chirality-reversible anomalous Hall effect in a magnetic kagome metal. Preprint at https://arxiv.org/abs/2409.01365 (2024).

[20] Ortiz, B. R. *et al*. New kagome prototype materials: discovery of $KV_3Sb_5$, $RbV_3Sb_5$, and $CsV_3Sb_5$. *Phys. Rev. Mater.* **3**, 094407 (2019).

[21] Yang, H. *et al*. Titanium-based kagome superconductor $CsTi_3Bi_5$ and topological states. Preprint at https://arxiv.org/abs/2209.03840 (2022).

[22] Werhahn, D. *et al*. The kagomé metals $RbTi_3Bi_5$ and $CsTi_3Bi_5$. *Z. Naturforsch. B* **77**, 757–764 (2022).

[23] Steiner, J. V. M. & Windsor, C. Theoretical and experimental studies on one-dimensional magnetic systems. *Adv. Phys.* **25**, 87 (1976).

[24] Bertini, B. *et al.* Finite-temperature transport in one-dimensional quantum lattice models. *Rev. Mod. Phys.* **93**, 025003 (2021).

[25] L. Šmejkal, J. Sinova & T. Jungwirth, Emerging research landscape of altermagnetism. *Phys. Rev. X* **12**, 040501 (2022).

[26] Jiang, Y.-X. *et al*. Unconventional chiral charge order in kagome superconductor $KV_3Sb_5$. *Nat. Mater.* **20**, 1353–1357 (2021).

[27] Wang, Z. *et al*. Electronic nature of chiral charge order in the kagome superconductor $CsV_3Sb_5$. *Phys. Rev. B* **104**, 075148 (2021).

[28] Liu, E. *et al*. Giant anomalous Hall effect in a ferromagnetic kagome-lattice semimetal. *Nat. Phys.* **14**, 1125–1131 (2018).

[29] Miyasato, T. *et al.* Crossover behavior of the anomalous Hall effect and anomalous Nernst effect in itinerant ferromagnets. *Phys. Rev. Lett.* **99**, 086602 (2007).

[30] Onoda, S., Sugimoto, N. & Nagaosa, N. Quantum transport theory of anomalous electric, thermoelectric, and thermal Hall effects in ferromagnets. *Phys. Rev. B* **77**, 165103 (2008).

[31] Manyala, N. *et al.* Large anomalous Hall effect in a silicon-based magnetic semiconductor. *Nat. Mater.* **3**, 255–262 (2004).

[32] Chen, T. *et al.* Anomalous transport due to Weyl fermions in the chiral antiferromagnets $Mn_3X$, $X$ = Sn, Ge. *Nat. Commun.* **12**, 572 (2021).

[33] Onoda, S., Sugimoto, N. & Nagaosa, N. Intrinsic versus extrinsic anomalous Hall effect in ferromagnets. *Phys. Rev. Lett.* **97**, 126602 (2006).

[34] Shen, J. *et al*., Intrinsically enhanced anomalous Hall conductivity and Hall angle in Sb-doped magnetic Weyl semimetal $Co_3Sn_2S_2$. *APL Mater.* **10**, 090705 (2022).

[35] P. Fazekas, Lecture Notes on Electron Correlation and Magnetism. World Scientific Publishing, Singapore 1999.

[36] Blöchl, P. E. Projector augmented-wave method. *Phys. Rev. B* **50**, 17953 (1994).

[37] Kresse, G. & Joubert, D. From ultrasoft pseudopotentials to the projector augmented-wave method. *Phys. Rev. B* **59**, 1758 (1999).

[38] Kresse, G. & Furthmüller, J. Efficiency of *ab*-initio total energy calculations for metals and semiconductors using a plane-wave basis set. *Comput. Mater. Sci.* **6**, 15–50 (1996).

[39] Kresse, G. & Furthmüller, J. Efficient iterative schemes for *ab initio* total-energy calculations using a plane-wave basis set. *Phys. Rev. B* **54**, 11169–11186 (1996).

[40] Perdew, J. P., Burke, K. & Ernzerhof, M. Generalized gradient approximation made simple.

*Phys. Rev. Lett.* **77**, 3865–3868 (1996).
[41] Küchler, R. *et al*. The world's smallest capacitive dilatometer, for high-resolution thermal expansion and magnetostriction in high magnetic fields. *Rev. Sci. Instrum.* **88**, 083903 (2017).

## Acknowledgments

This work was financially supported by the Deutsche Forschungsgemeinschaft (DFG) under SFB1143 (Project No. 247310070), the Würzburg-Dresden Cluster of Excellence on Complexity and Topology in Quantum Matter—ct.qmat (EXC 2147, Project No. 390858490), and Grant No. QUASTFOR5249-449872909. E.J.C. acknowledges the financial support from the Alexander von Humboldt Foundation. Z. K. Liu acknowledges the support from the National Natural Science Foundation of China (92365204, 12274298) and the National Key R&D program of China (Grant No. 2022YFA1604400/03). The SP-STM measurements (W.C. and W.W.) at Rutgers were supported by the Office of Basic Energy Sciences, Division of Materials Sciences and Engineering, U.S. Department of Energy under Award No. DE-SC0018153. Y.F.Y. acknowledges the support from the National Natural Science Foundation of China (12174429) and the National Key R&D program of China (Grant No. 2022YFA1402203). E.J.C. expresses gratitude to Yang Xu, Yu Pan, and Changjiang Yi for their insightful discussions, and to Dong Chen for assistance with thermal expansion measurements.

## Author Contributions

E.J.C. and Z.K.L. convinced the idea and designed the experiments. E.J.C. grew the single crystal, and conducted magnetization, electrical transport measurements, and transport data analysis. K.P.W., H.K.C., Y.W.L., M.X.W., Y.L.C. and Z.K.L. conducted ARPES experiments and analysis. W.Q.C., M.X.W., Z.K.L., and W.D.W. conducted STM measurements. S.M.N., H.X.T., and B.H.Y. performed DFT calculations. T.P.Y. helped to resolve the crystal structure. EJ.C., R.K. and W.S. measured heat capacity and thermal expansion. E.J.C., K.P.W., T.H.X., W.Q.C., and Z.K.L. contributed equally to this work. E.J.C., Y.F.Y., W.D.W., Z.K.L., and C.F. supervised the project. E.J.C., Z.K.L. and Y.F.Y. wrote the paper with input from all coauthors.

## Competing interests

The authors declare no competing interests.

## Additional Information

**Supplementary information** is available for this paper at the URL inserted when published.

**Correspondence** and requests for materials should be addressed to E.J.C. (Erjian.Cheng@cpfs.mpg.de), Y.F.Y. (yifeng@iphy.ac.cn), W.D.W. (wdwu@physics.rutgers.edu), Z.K.L. (liuzhk@shanghaitech.edu.cn), or C.F. (Claudia.Felser@cpfs.mpg.de).

# Figure 1

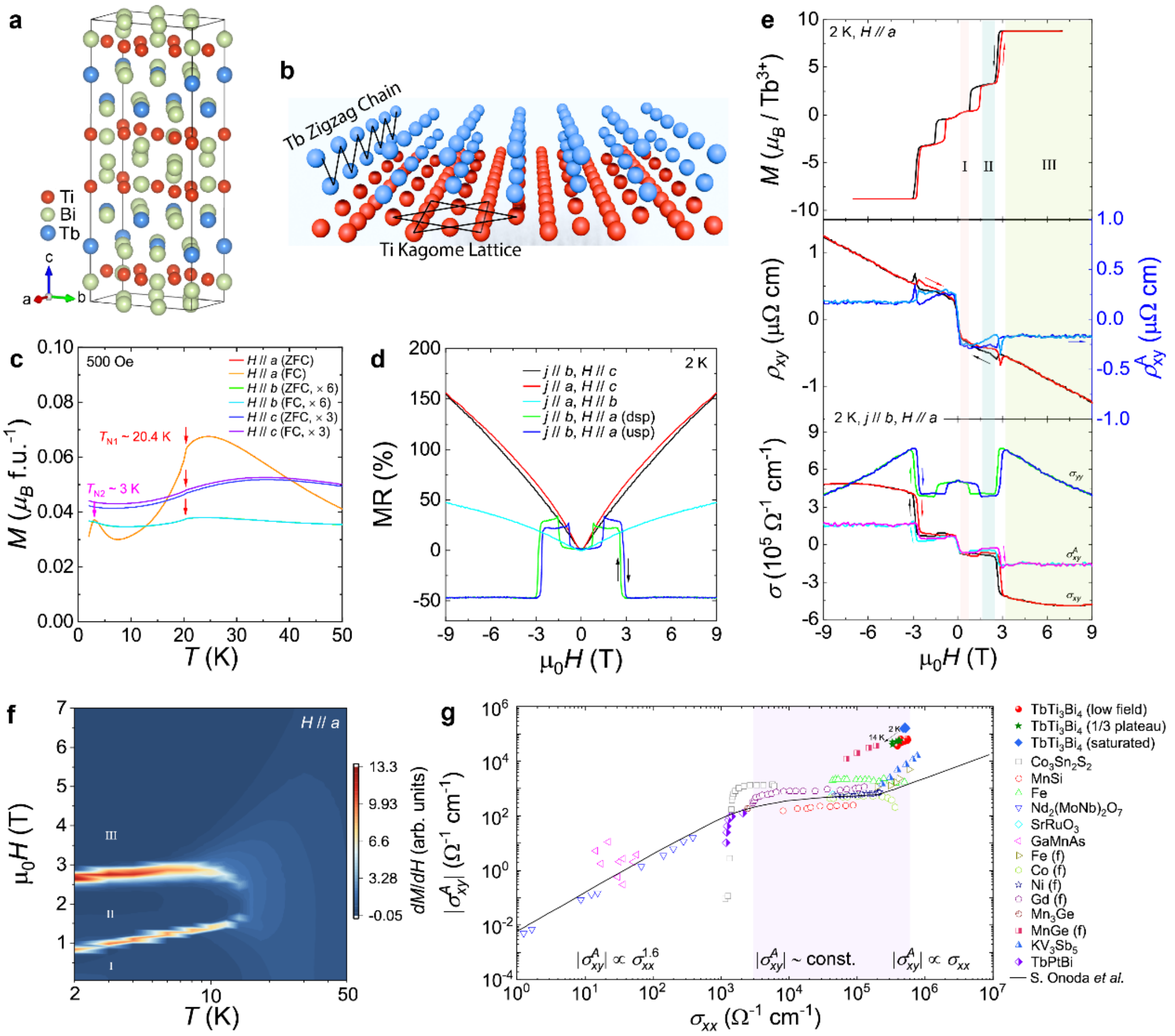

**Figure 1 | Interweaved crystal structure, magnetic characterization, and giant anomalous Hall conductivity in the magnetic kagome metal $TbTi_3Bi_4$.** (**a**) Side view of the $TbTi_3Bi_4$ crystal structure. (**b**) Schematic of the interwoven structure of the quasi one-dimensional (1D) Tb zigzag chain and Ti kagome layer in $TbTi_3Bi_4$. (**c**) Magnetization with magnetic field applied in different orientations during both zero-field cooling and field cooling processes. Two magnetic transitions located at 20.4 K ($T_{N1}$) and 3 K ($T_{N2}$), respectively, have been resolved. (**d**) Magnetoresistance (MR) at 2 K with different measurement configurations. When magnetic field is applied parallel to the *a* axis, MR show hysteresis behavior (dsp and usp represent field sweeping from 9 to -9 T and -9 to 9 T, respectively). (**e**) Upper panel: Field-dependent magnetization (MH) at 2 K for magnetic field applied parallel to the *a* axis. Magnetization exhibits pronounced hysteresis behavior, with a plateau at approximately 1/3 of the saturated magnetization of the saturated magnetization. Middle panel: Hall resistivity and anomalous Hall resistivity, with the linear conventional contribution subtracted. Lower panel: Calculated longitudinal and transverse conductivity, and anomalous Hall conductivity. The anomalous Hall effect contains three regimes, which are marked as I, II, and III, corresponding to the phase diagram obtained from magnetization (**f**). (**f**) Magnetic phase diagram of $TbTi_3Bi_4$ with

the magnetic field applied along the a axis. The background color indicates the magnitude of the magnetic susceptibility. **(g)** Full logarithmic plot of the absolute value of anomalous Hall conductivity ($|\sigma_{xy}^{A}|$) versus longitudinal conductivity ($\sigma_{xx}$) for $TbTi_3Bi_4$ with other reported various materials for comparison [1–3, 28–34]. The solid lines in three regimes represent $|\sigma_{\mathrm{xy}}^{\mathrm{A}}| \propto \sigma_{\mathrm{xx}}^{1.6}$, $|\sigma_{\mathrm{xy}}^{\mathrm{A}}|$ ~ const., and $|\sigma_{\mathrm{xy}}^{\mathrm{A}}| \propto \sigma_{\mathrm{xx}}$, for the dirty (localized hopping), intermediate (intrinsic), and clean (skew scattering) regimes (the shaded region), respectively [1,33]. The dashed lines represent the linear fit in the logarithmic coordinates.

# Figure 2

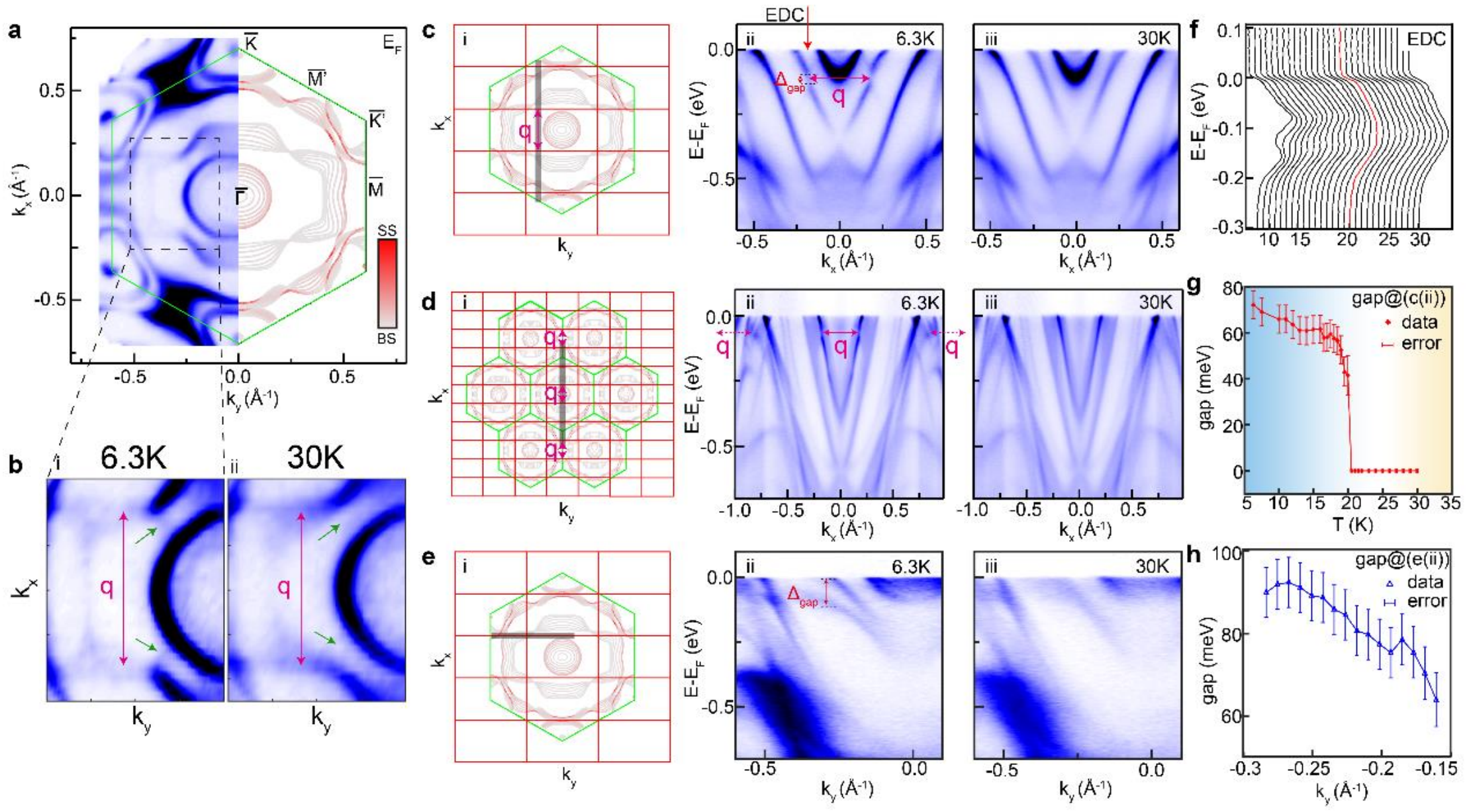

**Figure 2 | Electronic band folding and quasi-1D large hybridization gap of $TbTi_3Bi_4$ in the antiferromagnetic state.** (**a**) Constant energy contour at the Fermi energy (left panel) of $TbTi_3Bi_4$, plotted with the calculated band structure for the paramagnetic state (right panel). The band structure is measured with 46 eV photons with LH polarization, at $T \sim 6.3$ K. For calculations, the color bar represents the intensity of surface state projection. BS and SS represent bulk state and surface state, respectively. **b**(**i**) displays the zoom-in plots of the corresponding black dashed rectangular box in (**a**) with the spin-density-wave (SDW) vector labeled by **q**. Green arrows indicate the folded Fermi surfaces. **ii**) the data at 30 K for comparison. **c** (**i**) Schematic diagram illustrating the folding vector and the cut along the grey line in the Brillouin zone (BZ, the green lines). **ii**) and **iii**) show the intensity plots of energy band's dispersions along the cut along the grey line in (i) at 6.3 and 30 K, respectively. The dispersion is measured with 46 eV photons with LH polarization. At 6.3 K, the SDW gap $\Delta_{gap}$ in the AFM state can be clearly distinguished. **(d)** The same as **(c)**, but measured with 46 eV photons with LV polarization and along $k_y$=0. Red arrows indicate the signature of folded bands with Ti 3*d* orbitals and Bi 6*p* orbitals. (**e**) The same as (**c**), but measured along the quasi 1D Fermi surface. (**f**) Plot of the energy distribution curves (EDCs) labelled in **c** (**ii**) at various temperatures. The red line corresponds to 20K. (**g**) The extracted hybridization gap ($\Delta_{gap}$) size in **(f)** versus temperature. (**h**) The extracted hybridization gap ($\Delta_{gap}$) size in **e** (**ii**) versus $k_y$.

# Figure 3

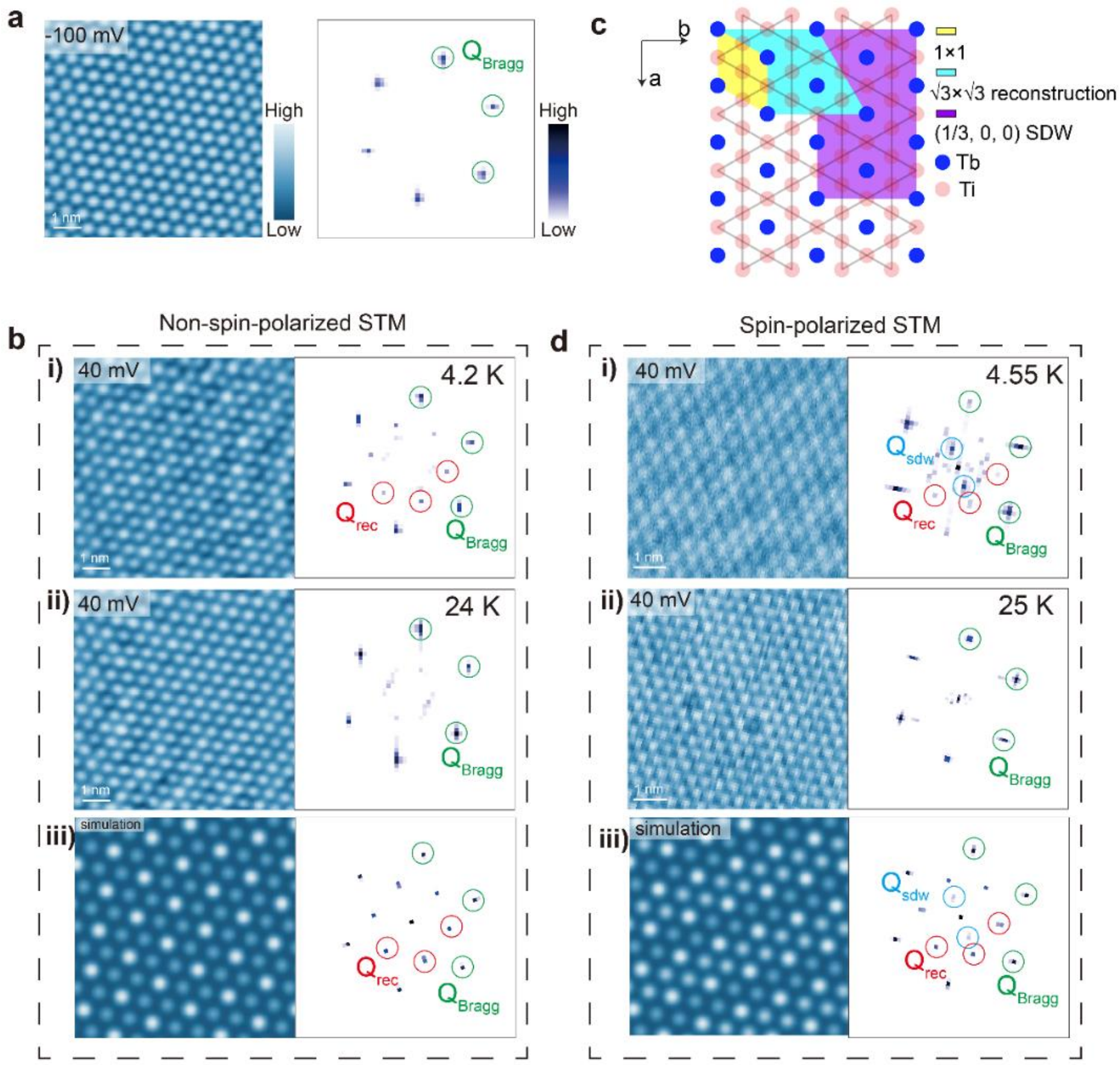

**Figure 3 | Observation of spin density wave and surface reconstruction of $TbTi_3Bi_4$ in the antiferromagnetic state from scanning tunneling microscopy.** (**a**) Topographic characterizations of the cleaved surface of $TbTi_3Bi_4$, showing one layer of Tb atoms. The right panel displays the fast Fourier transform (FFT) of the Tb lattice, clearly revealing hexagonal Bragg peaks. (**b**) LDOS map from non-spin-resolved STM obtained on an 8 nm × 8 nm area at -40 meV with FFT of the LDOS map (tunnelling setpoint: $I$ = 1 nA) at **i**) 4.2 K and **ii**) 24 K. **iii**) Simulated LDOS map and the corresponding FFT results. (**c**) A schematic diagram illustrating all observed reconstructions in $TbTi_3Bi_4$, with the yellow region representing the basis of a single unit cell. (**d**) LDOS map from spin-resolved STM obtained on an 8 nm × 8 nm area at -40 meV with FFT of the LDOS map (tunnelling setpoint: $I$ = 100 pA and 1.1 nA) at **i)** 4.55 K and **ii)** 25 K. **iii**) The simulated LDOS map and corresponding FFT results. The green/red/blue circles indicate the lattice Bragg ($Q_{Bragg}$)/reconstruction ($Q_{rec}$)/SDW ($Q_{SDW}$) peaks, respectively.

# Figure 4

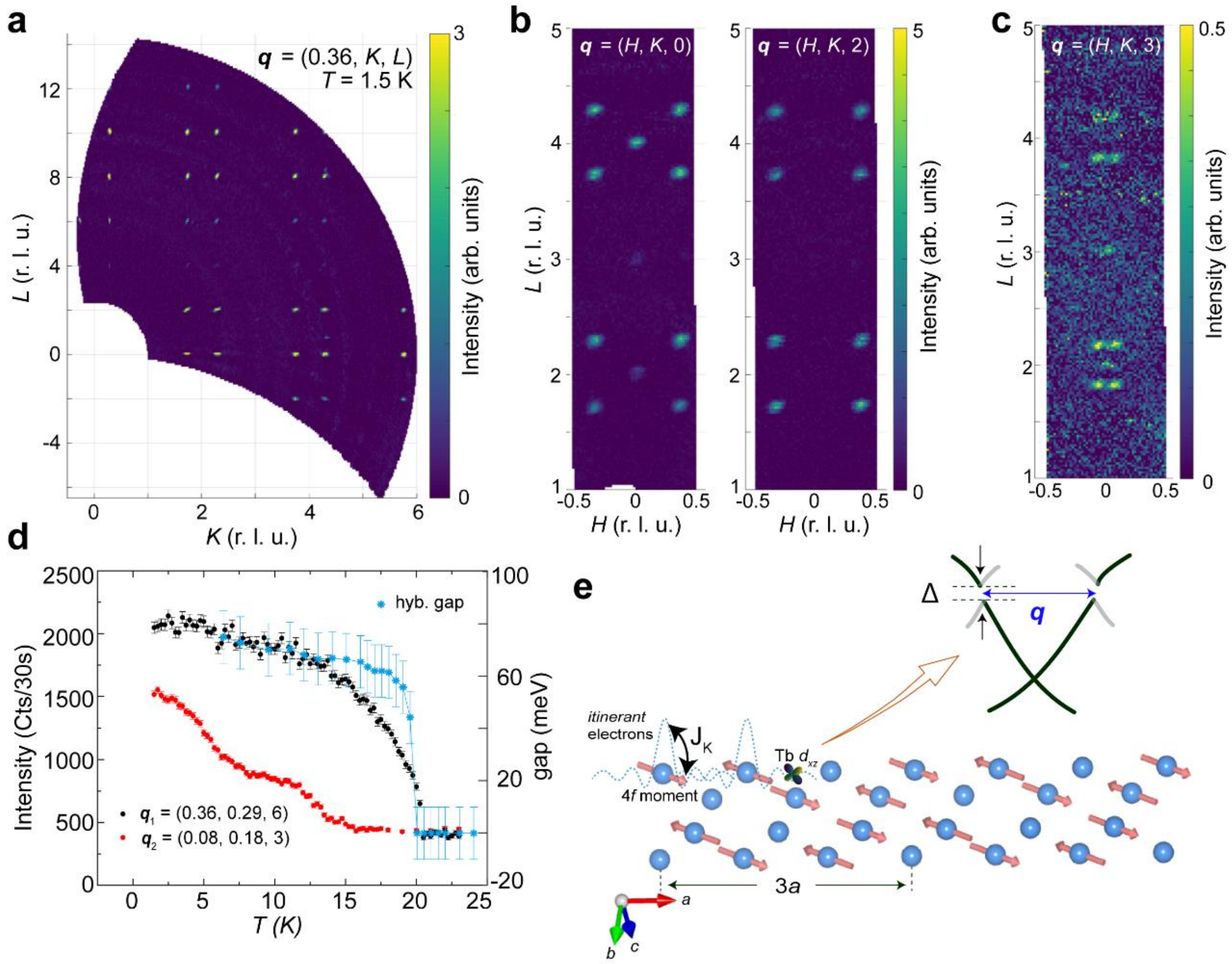

**Figure 4 | Magnetic structures of $TbTi_3Bi_4$ determined through neutron diffraction.** (**a**) $\boldsymbol{k_1}$ = (±0.36, ±0.29, 0) Magnetic peaks in the $H$ = 0.36 scattering plane. (**b**) $\mathbf{k_1}$ = (±0.36, ±0.29, 0) magnetic peaks in the $L$ = 0 (left panel) and $L$ = 2 (right panel) scattering planes. (**c**) $\boldsymbol{k_2}$ **=** (±0.08, ±0.18, 1) magnetic peaks observed in the $L$ = 3 scattering plane. (**d**) Temperature dependence of the intensity of both wavevectors ($\boldsymbol{q}_1$ and $\boldsymbol{q}_2$), overlapped with the temperature dependence hybridization gap plot as in Fig. 2g for comparison. (**e**) Schematic diagram illustrating the origin of the large electronic band hybridization gap due to the Kondo coupling between itinerant electron and localized magnetic moments. Ellipsoid spiral magnetic structures of the $\boldsymbol{k_1}$ magnetic order have been shown.

**Supplementary Information for**

# Spectroscopic origin of giant anomalous Hall effect in an interwoven magnetic kagome metal

Erjian Cheng[1,#,*], Kaipu Wang[2,#], Yiqing Hao[3,#], Wenqing Chen[4,#], Hengxin Tan[5,#], Zongkai Li[2,#], Meixiao Wang[2], Wenli Gao[6], Di Wu[6], Shuaishuai Sun[7], Tianping Ying[7], Simin Nie[8], Yiwei Li[9], Walter Schnelle[1], Houke Chen[10], Xingjiang Zhou[6,7], Ralf Koban[1], Yulin Chen[2,10], Binghai Yan[5], Yi-feng Yang[6,7,11,*], Weida Wu[4,*], Zhongkai Liu[2,*] and Claudia Felser[1,*]

[1]*Max Planck Institute for Chemical Physics of Solids, 01187 Dresden, Germany*

[2]*School of Physical Science and Technology, ShanghaiTech Laboratory for Topological Physics, ShanghaiTech University, 201210 Shanghai, China*

[3]*Neutron Scattering Division, Oak Ridge National Laboratory, P.0. Box 2008, Oak Ridge, Tennessee, 37831, USA*

[4]*Department of Physics & Astronomy, Rutgers University, Piscataway, NJ 08854, USA*

[5]*Department of Condensed Matter Physics, Weizmann Institute of Science, Rehovot 7610001, Israel*

[6]*University of Chinese Academy of Sciences, Chinese Academy of Sciences, 100190 Beijing, China*

[7]*Beijing National Laboratory for Condensed Matter Physics, Institute of Physics, Chinese Academy of Sciences, Beijing 100190, China*

[8]*Department of Mechanical Engineering, Stanford University, 94305 Stanford, California, USA*

[9]*Institute for Advanced Studies (IAS), Wuhan University, 430072 Wuhan, China*

[10]*Department of Physics, University of Oxford, Oxford OX1 3PU, UK*

[11]*Songshan Lake Materials Laboratory, Dongguan, Guangdong 523808, China*

**This PDF file includes:**

**Supplementary Note 12:** Thermal expansion and magnetostriction experiments

**Supplementary Figure 1 to 29**

## Supplementary Note 1: Schematic of kagome systems with charge density and spin density waves and their driving mechanisms

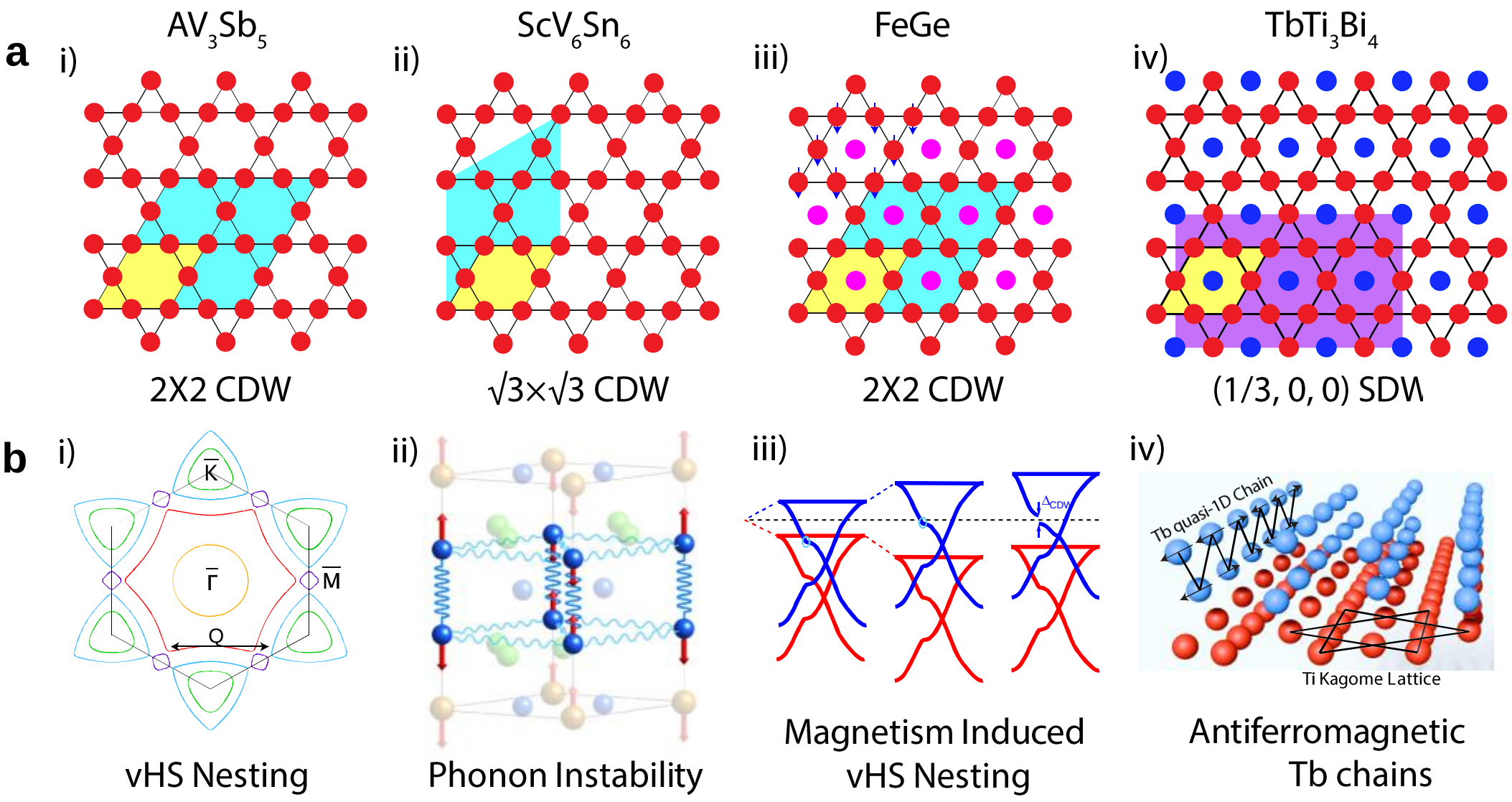

**Supplementary Figure 1 | Schematic of kagome systems with (a) charge density wave/spin density waves and (b) their driving mechanisms.** (**i**) For the $AV_3Sb_5$ (A = K, Rb, Cs) systems, a van Hove singularity (vHS) nesting scenario has been proposed for the charge-density-wave (CDW) instability [1–4]. (**ii**) For $ScV_6Sn_6$, the CDW may arise from phonon instability [5]. (**iii**) For FeGe, the CDW might come from magnetism-induced vHS nesting [6,7]. (**iv**) The observation of SDW order in $TbTi_3Bi_4$.

## Supplementary Note 2: Heat capacity and the inverse magnetic susceptibility

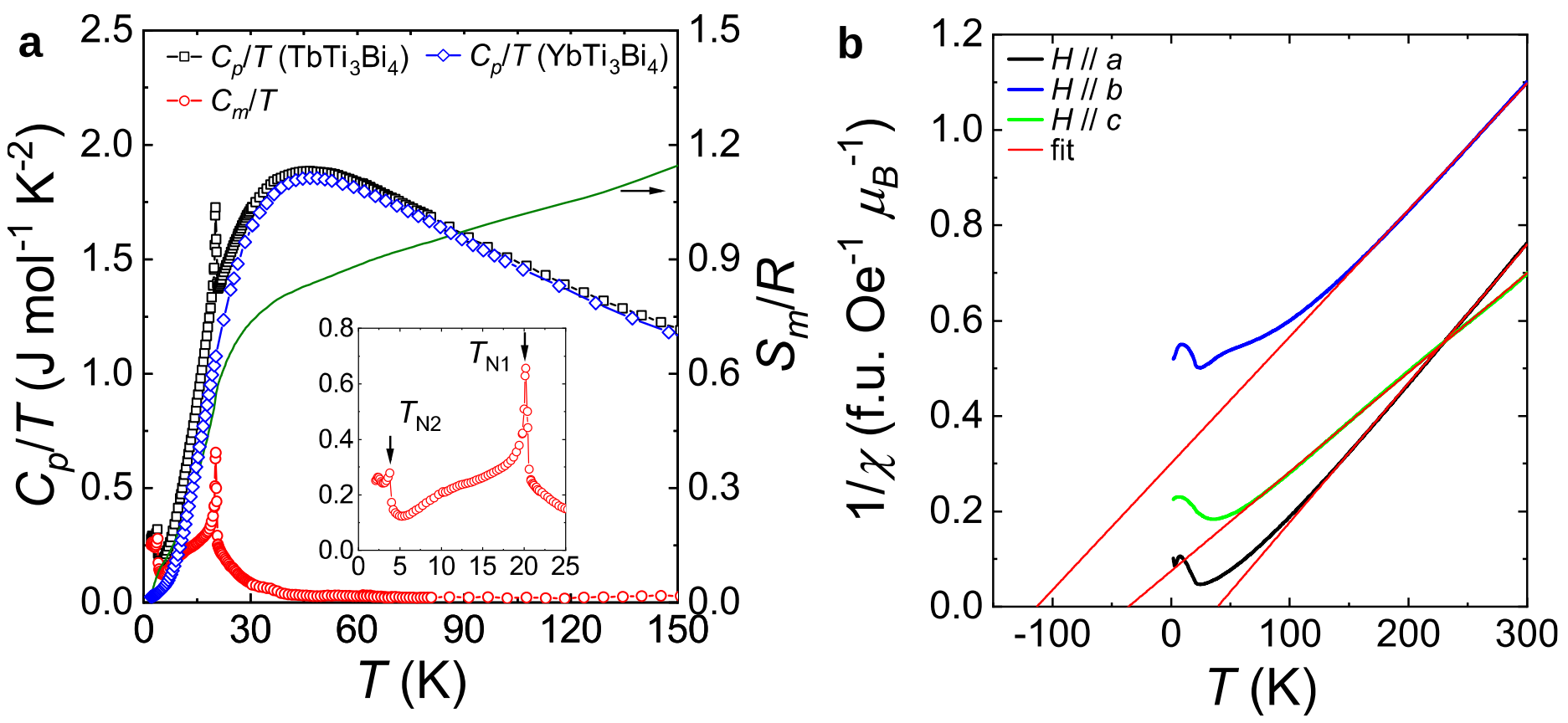

**Supplementary Figure 2 | Heat capacity and the inverse magnetic susceptibility.** (**a**) Heat capacity of $TbTi_3Bi_4$ in zero field applied along *c* axis. Nonmagnetic counterpart $YbTi_3Bi_4$ has been also plotted for subtracting the magnetic contributions. Two magnetic transitions at 20.4 K ($T_{N1}$) and 3 K ($T_{N2}$) are clearly visible. The solid line represents magnetic entropy. The entropy above $T_N$

is around $R\ln 2 = 0.69\ R$, indicating that the magnetic ordering comprises two $m_J$ levels, could be a doublet ground state. These levels are pretty well separated in energy, since the magnetic entropy increases only slowly with temperature ($R\ln3 = R \times 1.1$) is only seen above 130 K. Thus, the next $m_J$ state energy (difference to the lowest level) is only in this. (**b**) The inverse magnetic susceptibility for different orientations. For *a* axis, it shows ferromagnetic coupling with a Weiss temperature of 40 K, while antiferromagnetic coupling with values of -113 and -35 K for *b* and *c* axis, respectively.

## Supplementary Note 3: Electrical transport and anomalous Hall effect

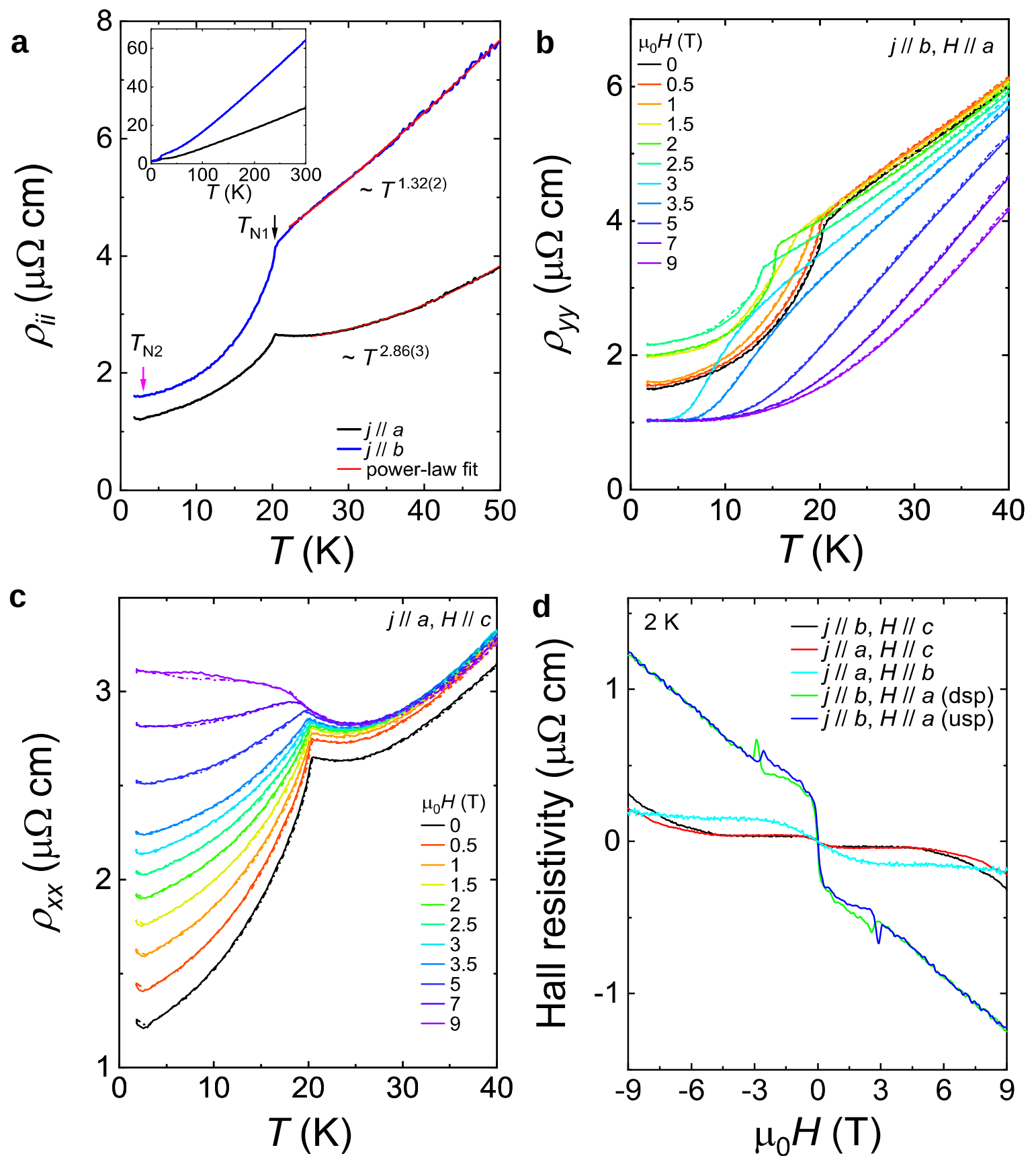

**Supplementary Figure 3 | Electrical transport properties and magnetic anisotropy of $TbTi_3Bi_4$.** **(a)** Low-temperature longitudinal resistivity with current applied along different orientations. Inset shows the longitudinal resistivity at high temperature. Red solid lines represent the fit to the data using a power law, $\rho_{ii} = \rho_0 + AT^n$, where $\rho_0$ is the residual resistivity, $A$ and $n$ are fitting parameters. For the current applied along the *a* axis, the fit gives a $T^{2.86(3)}$ dependence, while a $T^{1.32(2)}$ dependence for the current applied along the *b* axis, indicating strong scattering for this orientation. (**b**) Resistivity at several fixed fields with current and field applied along the *b* axis and *a* axis, respectively. (**c**) Resistivity at several fixed fields with current and field applied along the *a* axis and

*c* axis, respectively. The solid and dashed lines represent warming and cooling process, respectively. When temperature approaches $T_{N1}$, the resistivity slightly increases, suggestive of a gap opening, in line with the observation of the quasi-1D hybridization gap formation in ARPES experiments. Under an applied magnetic field, the resistivity exhibits different behaviors depending on the field orientation, showing a nonmonotonic evolution only when the field is aligned along the *a* axis. (**d**) Hall resistivity at 2 K with different measurement configurations. When magnetic field is applied parallel to the *a* axis, Hall resistivity shows distinct hysteresis behavior (dsp and usp represent field sweeping from 9 to -9 T and -9 to 9 T, respectively), consistent with magnetization measurements.

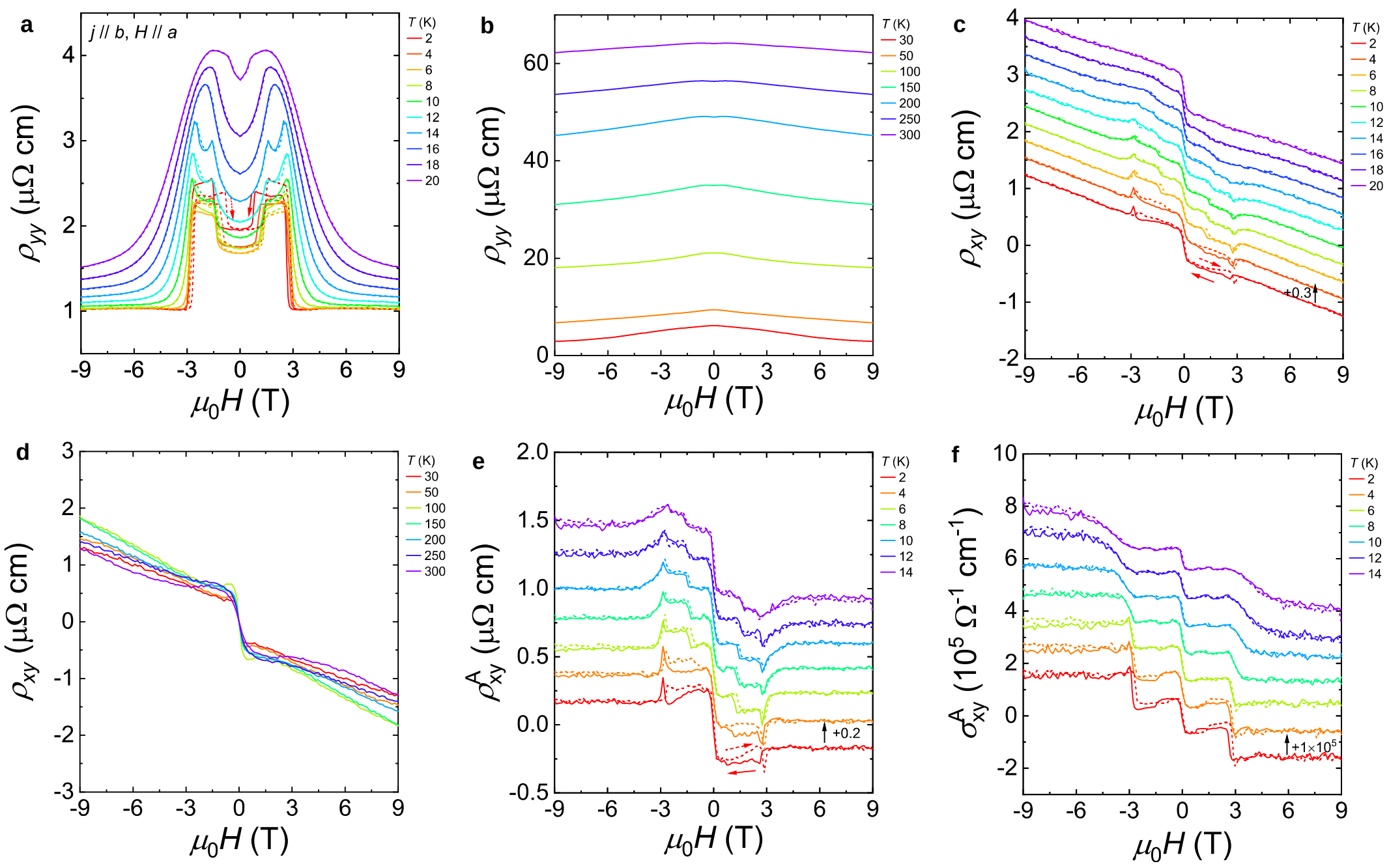

**Supplementary Figure 4 | Transport data and anomalous Hall effect with magnetic field and current applied along the *a* and *b* axis, respectively.** (**a,b**) Longitudinal resistivity at various temperatures with magnetic field applied along the *a* axis. Obviously, negative magnetoresistance (MR) exists up to 300 K. The sold and dashed arrows represent field sweeping from 9 to -9 T, and -9 to 9 T, respectively. Negative longitudinal MR is typically attributed to several factors [8–10]: (i) inhomogeneous current distribution or current jetting effect, (ii) weak localization effects, (iii) the chiral anomaly of Weyl fermions, and (iv) field-induced suppression of scattering from local moments or magnetic impurities. Given the measurement configurations and the broad temperature dependence of the negative MR in our experiments, scenarios (i), (ii), and (iii) can be excluded. We therefore suggest that magnetic fluctuations in the paramagnetic state are responsible. However, the observation of negative MR extending to room temperature in $TbTi_3Bi_4$ is very rare, indicating that it is the easy damping of the spin fluctuation components perpendicular to the long-range ordering vector (*viz*, the *a* axis) when applying a field along this direction. (**c,d**) Transverse resistivity ($\rho_{xy}$) at various temperatures with magnetic field applied along the *a* axis. For (**c**), the data is shifted by

0.3 μΩ cm for clarity. (**e,f**) The obtained anomalous Hall resistivity ($\rho_{xy}^{A}$) and anomalous Hall conductivity ($\sigma_{xy}^{A}$) in the magnetic states, respectively.

## Supplementary Note 4: The core-level spectrum of $TbTi_3Bi_4$

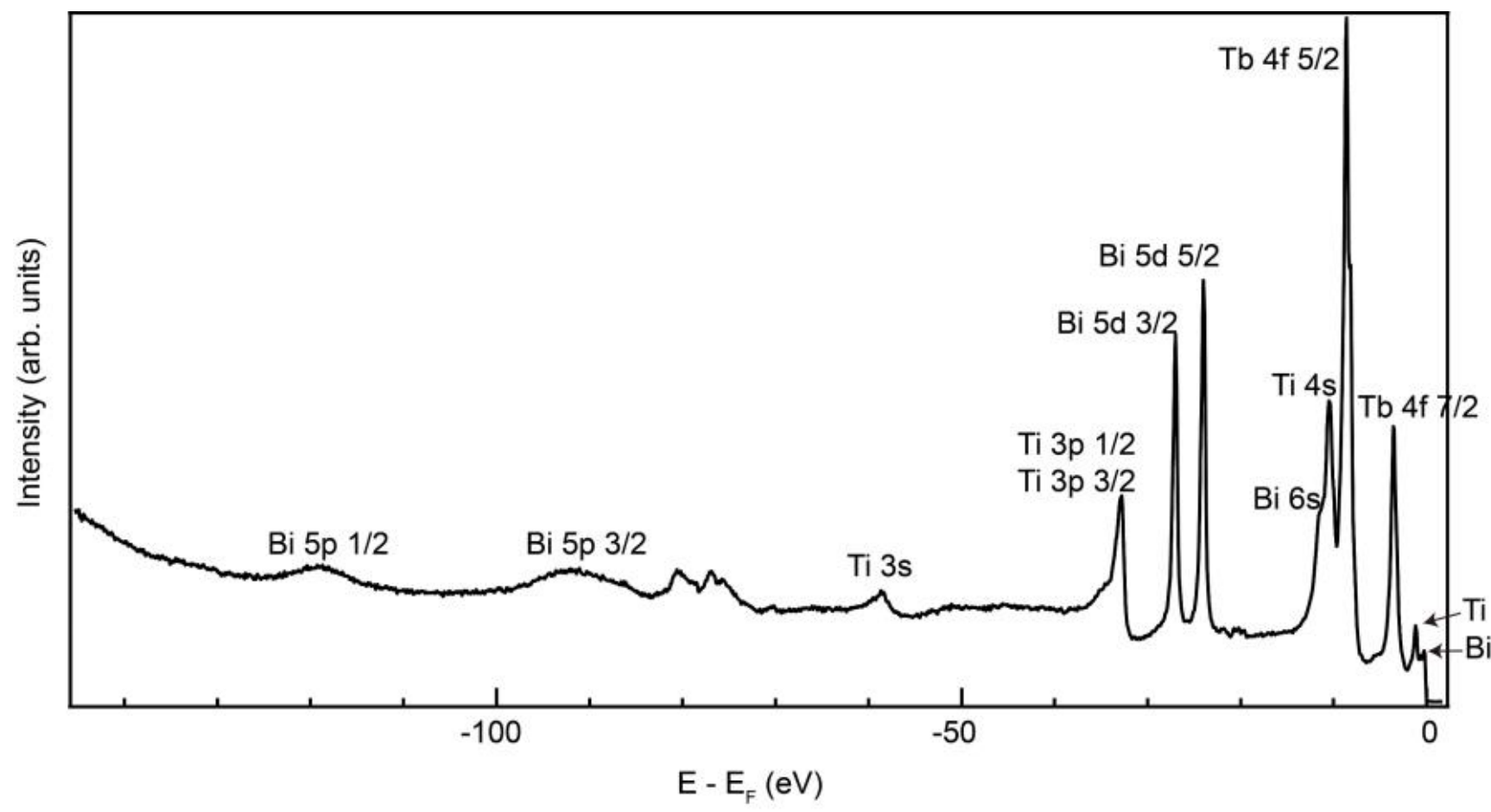

**Supplementary Figure 5 | The core-level spectrum of $TbTi_3Bi_4$ with the characteristic elemental peaks labelled.**

## Supplementary Note 5: Additional data on the electronic structure of $TbTi_3Bi_4$

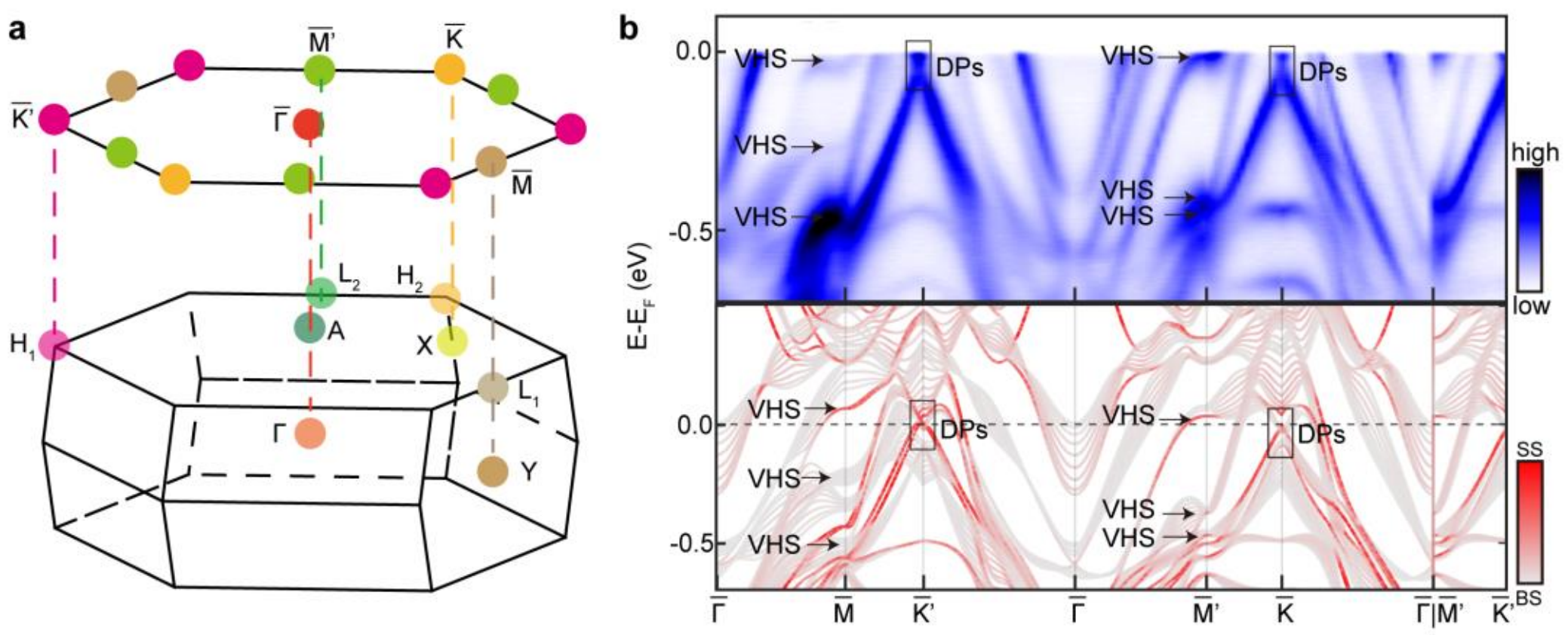

**Supplementary Figure 6 | Overall electronic structure of $TbTi_3Bi_4$. (a)** Illustration of the 3D BZ of $TbTi_3Bi_4$, and the corresponding (001) surface BZ with the high symmetry points marked. **(b)** The electronic structure of $TbTi_3Bi_4$ in the $k_z$ = 0 plane, measured with 46eV, LH polarization, at 30K.

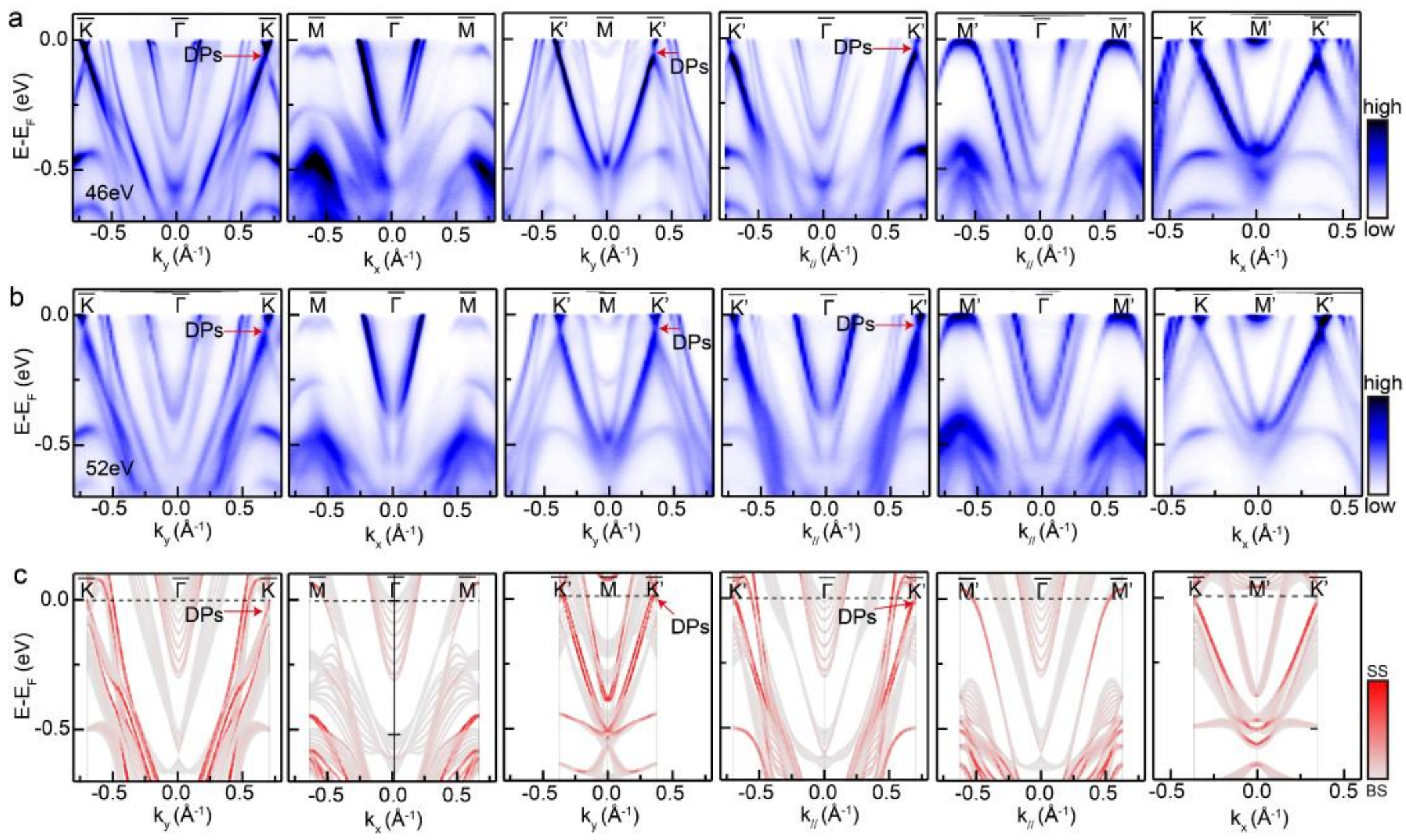

**Supplementary Figure 7 | The electronic energy bands along the high symmetry directions of $TbTi_3Bi_4$ measured with LH polarization at 6.3 K.** (**a**) The ARPES intensity plots of $TbTi_3Bi_4$ along various high symmetry directions with the DPs marked. The bandstructure was measured with photon energy 46 eV (near Γ plane). (**b**) The same as (**a**), but measured with photon energy 52 eV (near Z plane). The electronic structures near Γ and Z plane are almost identical, indicating significant $k_z$-broading effect. (**c**) The calculated energy band projection along various high symmetry directions. The measured energy bands are consistent with the calculated results.

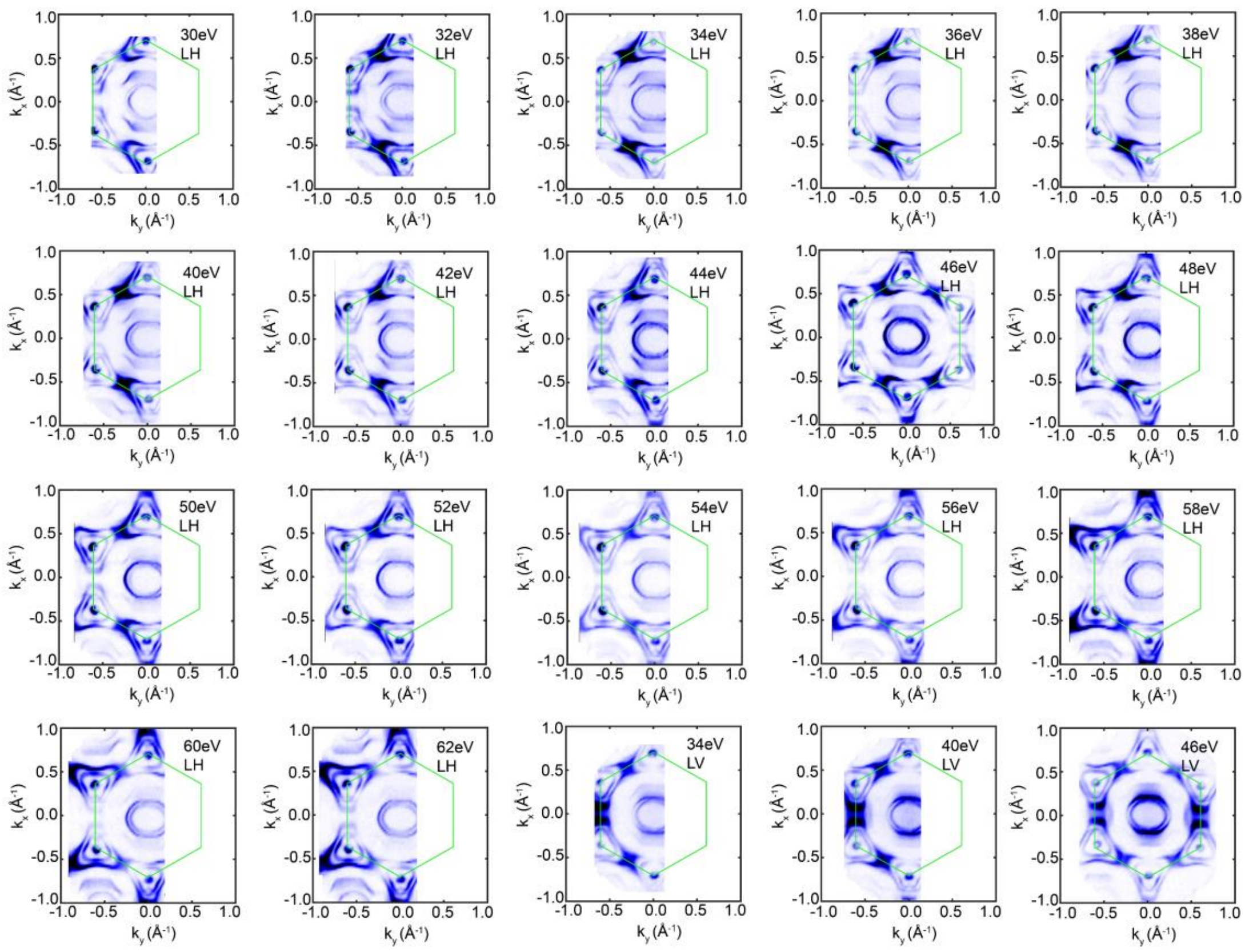

**Supplementary Figure 8 | The Fermi surfaces of $TbTi_3Bi_4$ measured with different incident photons and different polarizations at 6.3 K. The energy integration window is from -0.005 eV to $E_F$.**

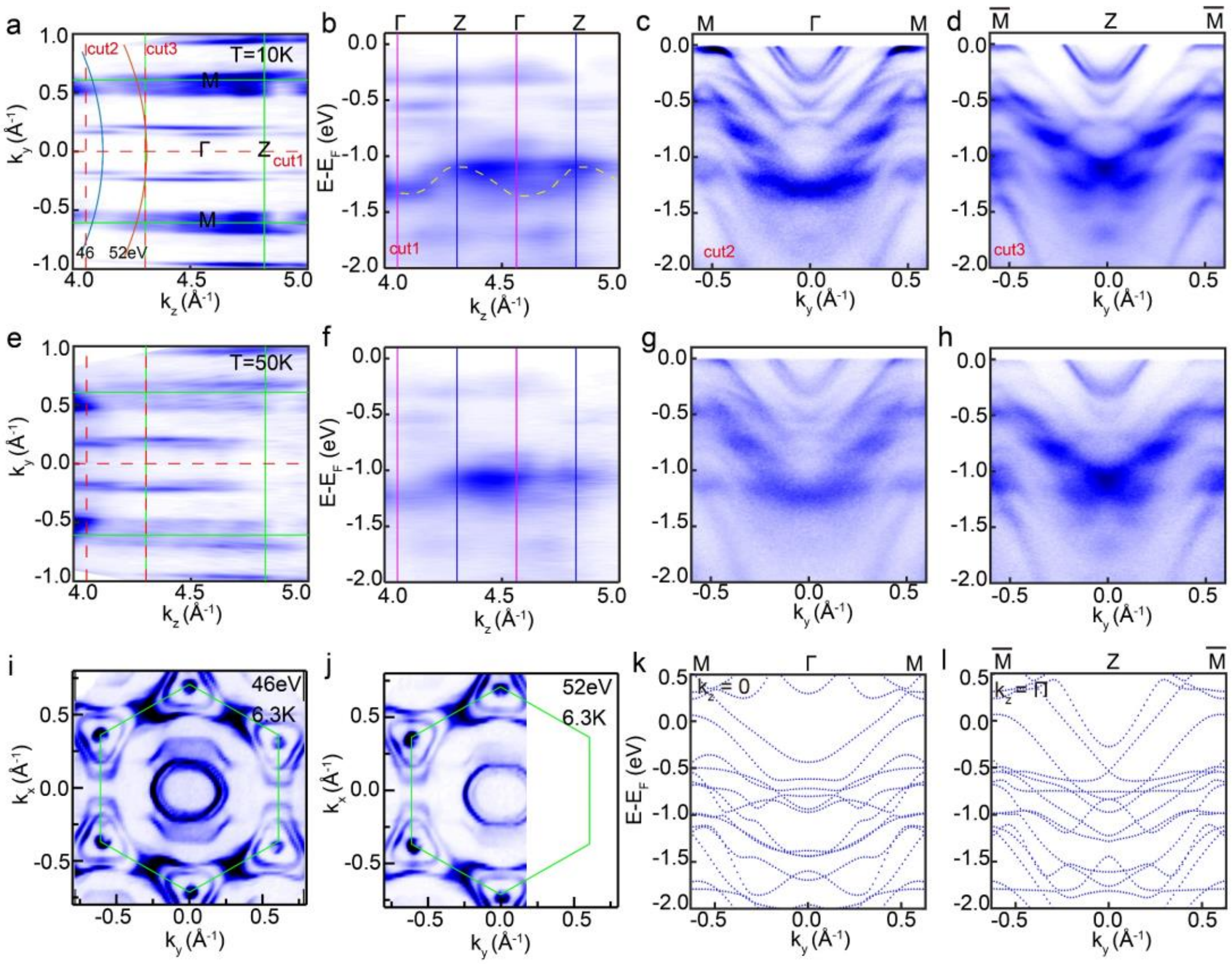

**Supplementary Figure 9 | The $k_z$ evolution of the electronic structure in $TbTi_3Bi_4$.** (**a**) The Fermi surface map as a function of $k_y$ and $k_z$ along the $\overline{M} - \overline{\Gamma} - \overline{M}$ direction using photon energies from 40 to 80 eV, with photon energies 46 eV (near Γ) and 52 eV (near Z) marked by the blue and orange curves, respectively. The green lines represent the 3D BZ. We could observe that the Fermi surface located around ± 0.2 $Å^{-1}$ show significant periodic intensity variation through varying photon energy. (**b**–**d**) The plot of the $k_z$ versus E dispersion along the red dashed lines in (**a**). The dispersion marked by the yellow dashed curve in (**b**) shows periodic changes along $k_z$ direction. (**c**) and (**d**) are along the high symmetry direction $M - \Gamma - M$ and $\overline{M} - Z - \overline{M}$ corresponding to the $k_z = 0$ and π planes respectively. (**e**–**h**) The same as (**a**–**d**) but measured at 50 K. (**i**, **j**) The Fermi surface map measured as a function of $k_x$ and $k_y$ at $k_z = 0$ and π plane at 6.3 K, respectively. (**k**, **l**) The calculated energy bands along the high symmetry direction $M - \Gamma - M$ and $\overline{M} - Z - \overline{M}$ at $k_z = 0$ and π planes respectively.

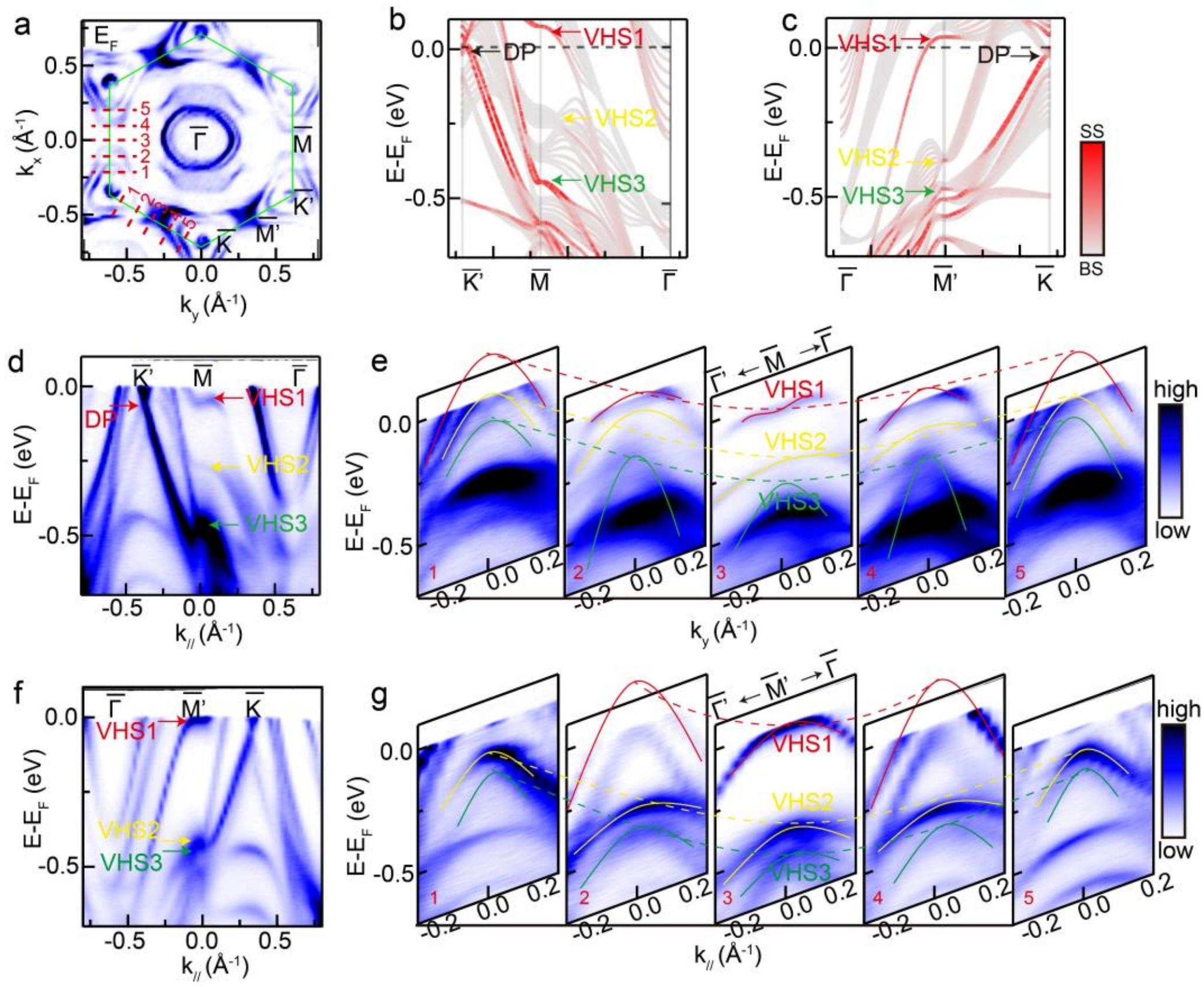

**Supplementary Figure 10 | The vHSs of $TbTi_3Bi_4$ measured with photon energy 46 eV at 6.3 K, with LH polarization.** (**a**) The Fermi surface of $TbTi_3Bi_4$ measured with photon energy 46 eV at 6.3 K, with LH polarization. (**b**, **c**) The projected band structure along the $\bar{K}' - \bar{M} - \bar{\Gamma}$ and $\bar{\Gamma} - \overline{M'} - \bar{K}$ directions with the VHSs and DP marked. (**d**, **f**) The ARPES intensity of the bandstructure along the $\bar{K}' - \bar{M} - \bar{\Gamma}$ and $\bar{\Gamma} - \overline{M'} - \bar{K}$ directions, with three vHSs marked by red, yellow, green arrows, respectively. The VHSs are consistent with those in (**b**, **c**). (**e**, **g**) The cut 1 – 5 plot the ARPES spectra measured perpendicular to the $\bar{K}' - \bar{M} - \bar{K}'$ and $\bar{K}' - \overline{M'} - \bar{K}$ directions, with the cut 3 crossing the M/M' points (see the red dashed lines in **a** for the exact momentum positions of cuts 1 – 5). We identified three coexisting VHSs in $TbTi_3Bi_4$ as marked by red, yellow, and green guidelines, respectively.

## Supplementary Note 6: The calculated bandstructure with atomic orbital projection of $TbTi_3Bi_4$

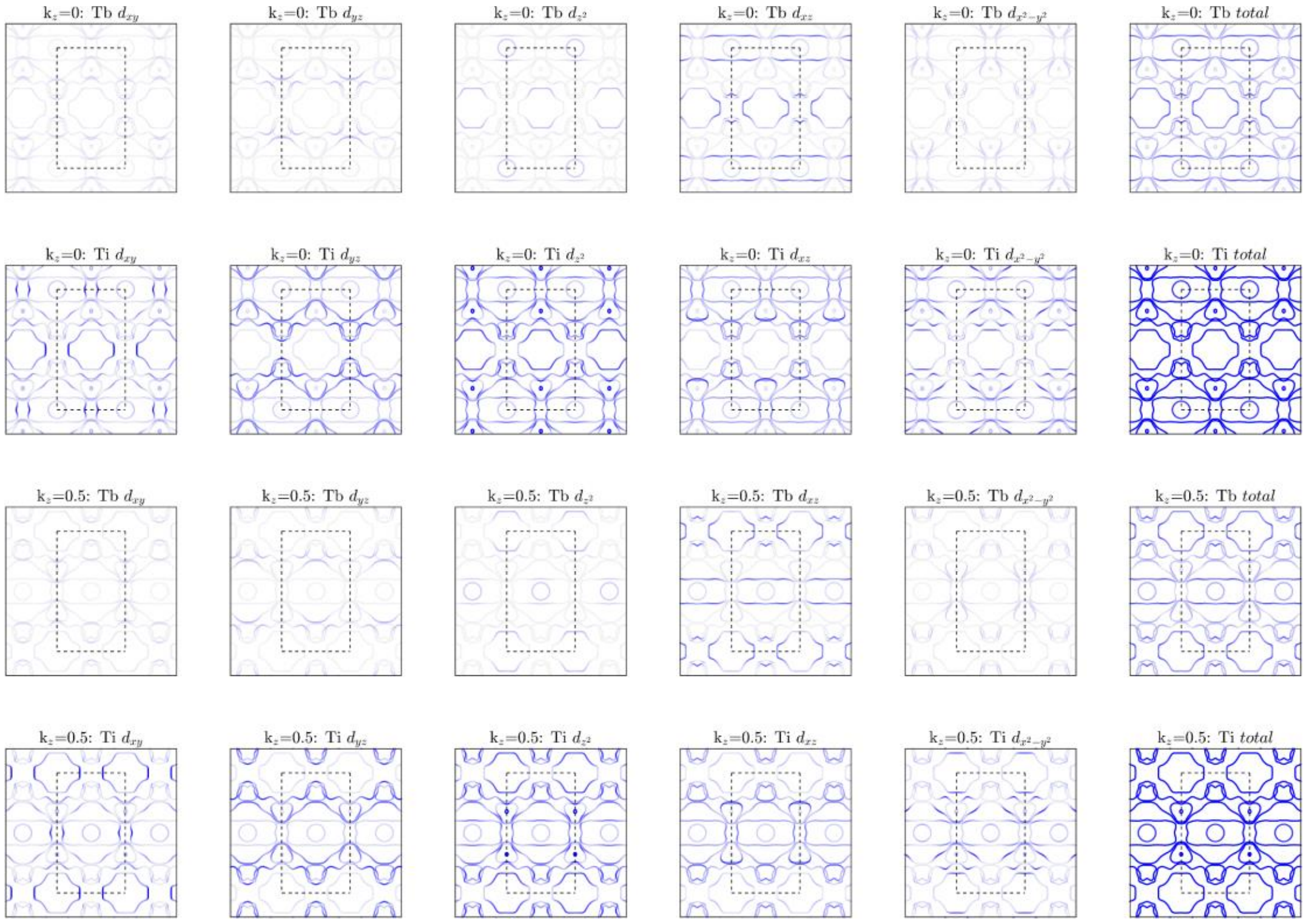

**Supplementary Figure 11 | The calculated Fermi surfaces with atomic orbital projection of $TbTi_3Bi_4$.** Based on the calculated band structure with orbital projection, one can see the folded quasi-1D bands comes from Tb $d_{xz}$ orbitals.

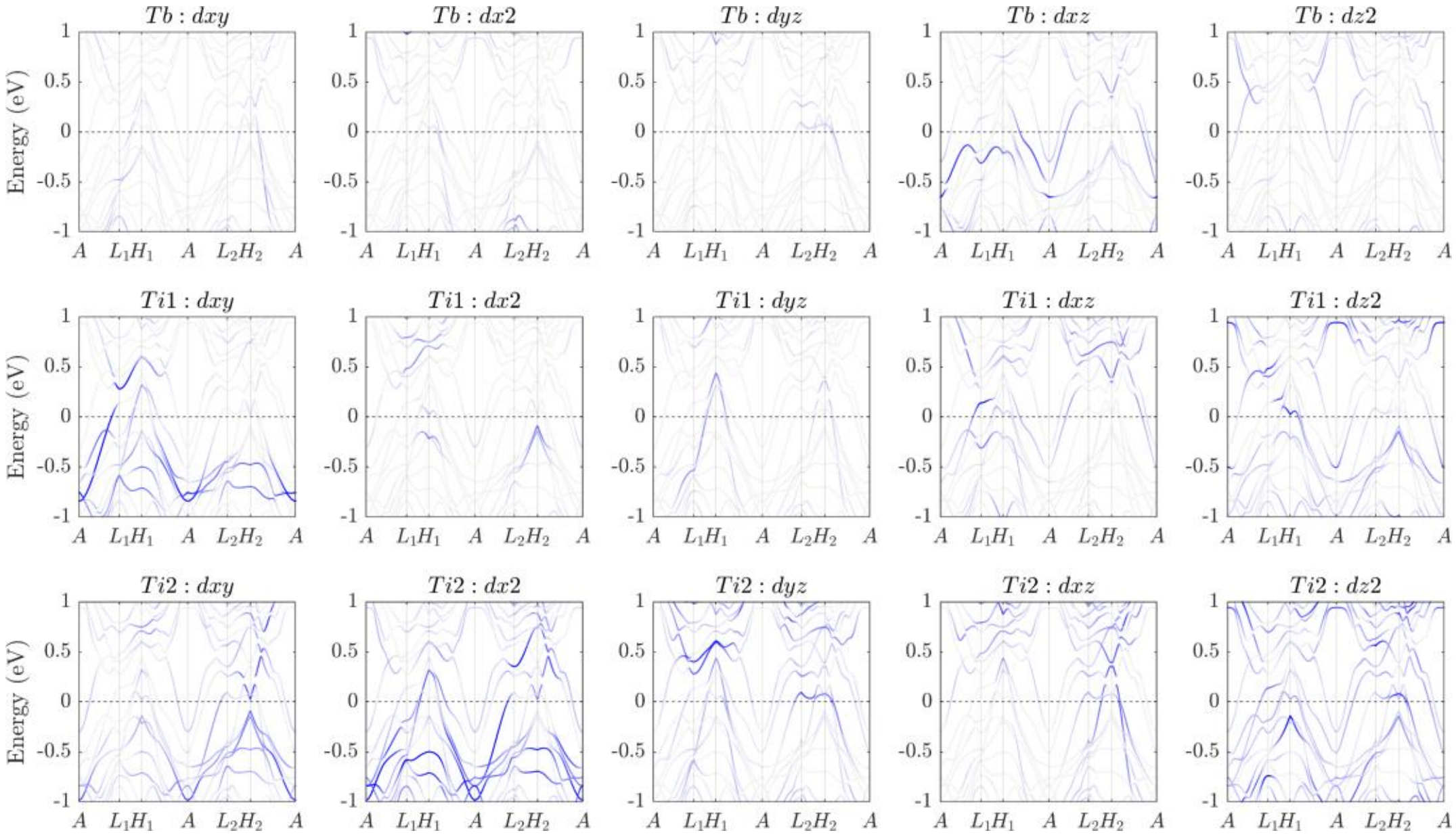

**Supplementary Figure 12 | The calculated electronic structure of $TbTi_3Bi_4$ along the high symmetry directions in the $k_z$ = π plane, with the contribution of different orbits highlighted.**

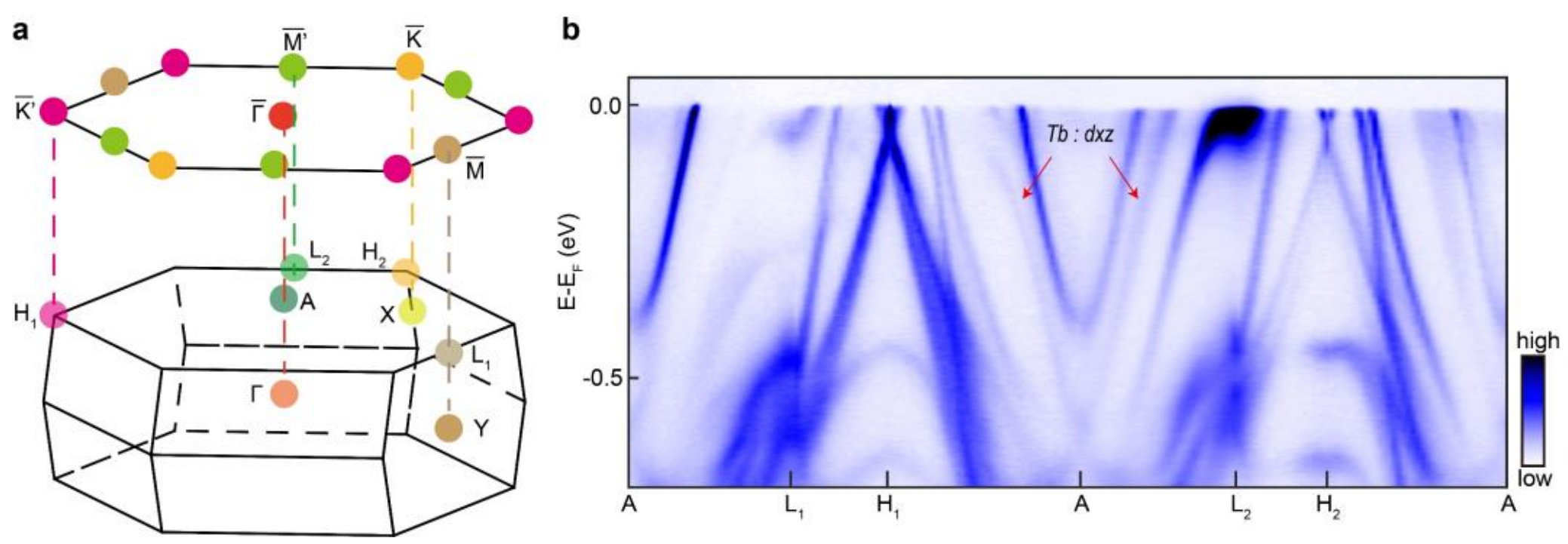

**Supplementary Figure 13 | Identification of the Tb $d_{xz}$ band. (a)** Illustration of the 3D BZ of $TbTi_3Bi_4$, and the corresponding (001) surface BZ with the high symmetry points marked. **(b)** The electronic structure of $TbTi_3Bi_4$ in the $k_z = \pi$ plane, measured with 36 eV, LH polarization, at 6.3K. Compared with the electronic bands in Supplementary Figure 12, the Tb $d_{xz}$ bands are marked by red arrows.

## Supplementary Note 7: The evolution of folded bands in the momentum space

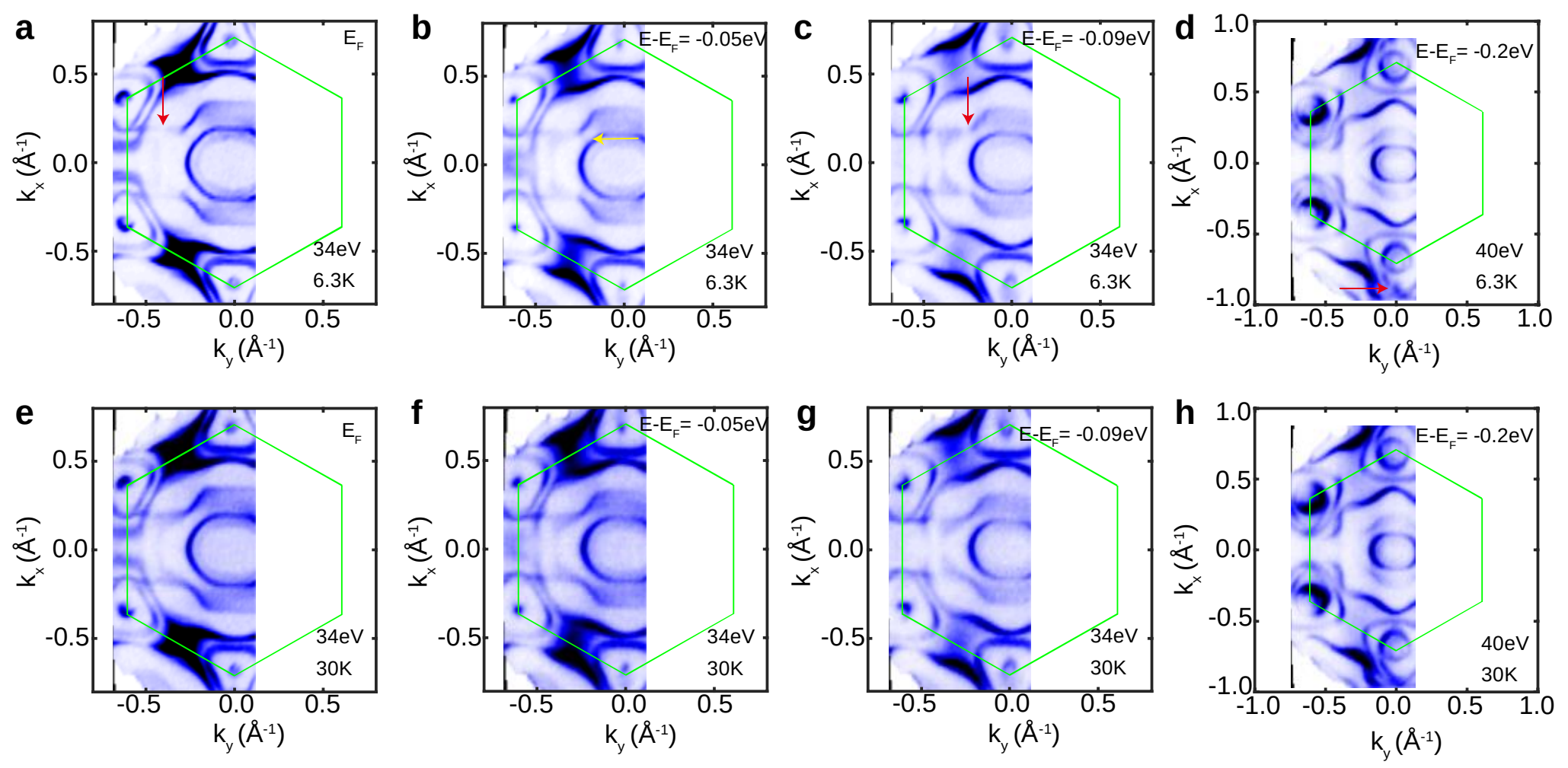

**Supplementary Figure 14 | The constant energy surfaces of $TbTi_3Bi_4$.** (**a**–**d**) The constant energy surfaces of $TbTi_3Bi_4$ at $E_F$ and 0.05, 0.09, 0.2 eV below $E_F$, respectively, measured at 6.3 K, with photon 34 eV for **a**–**c** and 40 eV for (**d**). (**e**–**h**) The same as (**a**–**d**), but measured at 30 K. Comparing the constant energy surfaces in AFM and PM states, one can see that the suppression of the spectral weight (marked by red arrows in **a**, **c**, **d**) and folded Fermi surface (marked by yellow arrow in **b**) emerge in AFM state. According to the analysis in the main text, the small gaps in **a**, **c** correspond to the hybridization gap from band folding of the Tb 5*d* bands caused by band folding, while the gap in **d** corresponds to the band folding from the band folding of the Ti 3*d* bands.

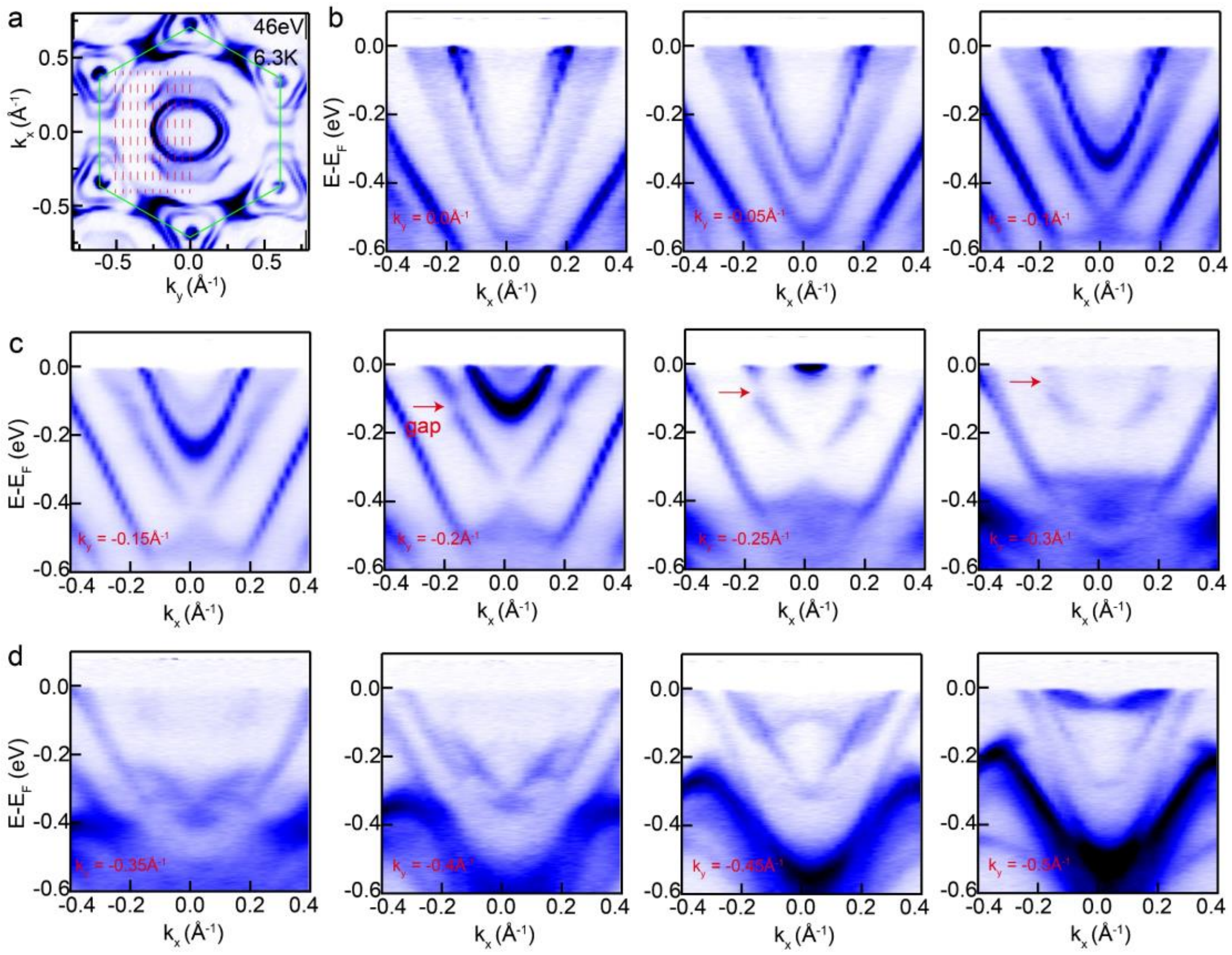

**Supplementary Figure 15 | The hybridization gap evolution along the $k_y$ direction in AFM state measured with photon energy 46 eV, LH polarization, at 6.3 K.** (**a**) The Fermi surface map of $TbTi_3Bi_4$ measured with photon energy 46 eV, LH polarization, at 6.3 K. (**b–d**) The ARPES intensity plots along the red dashed line in (**a**), from right to left. The hybridization gap is significant with $k_y$ between -0.2 to -0.3 Å$^{-1}$, which is marked by red arrow.

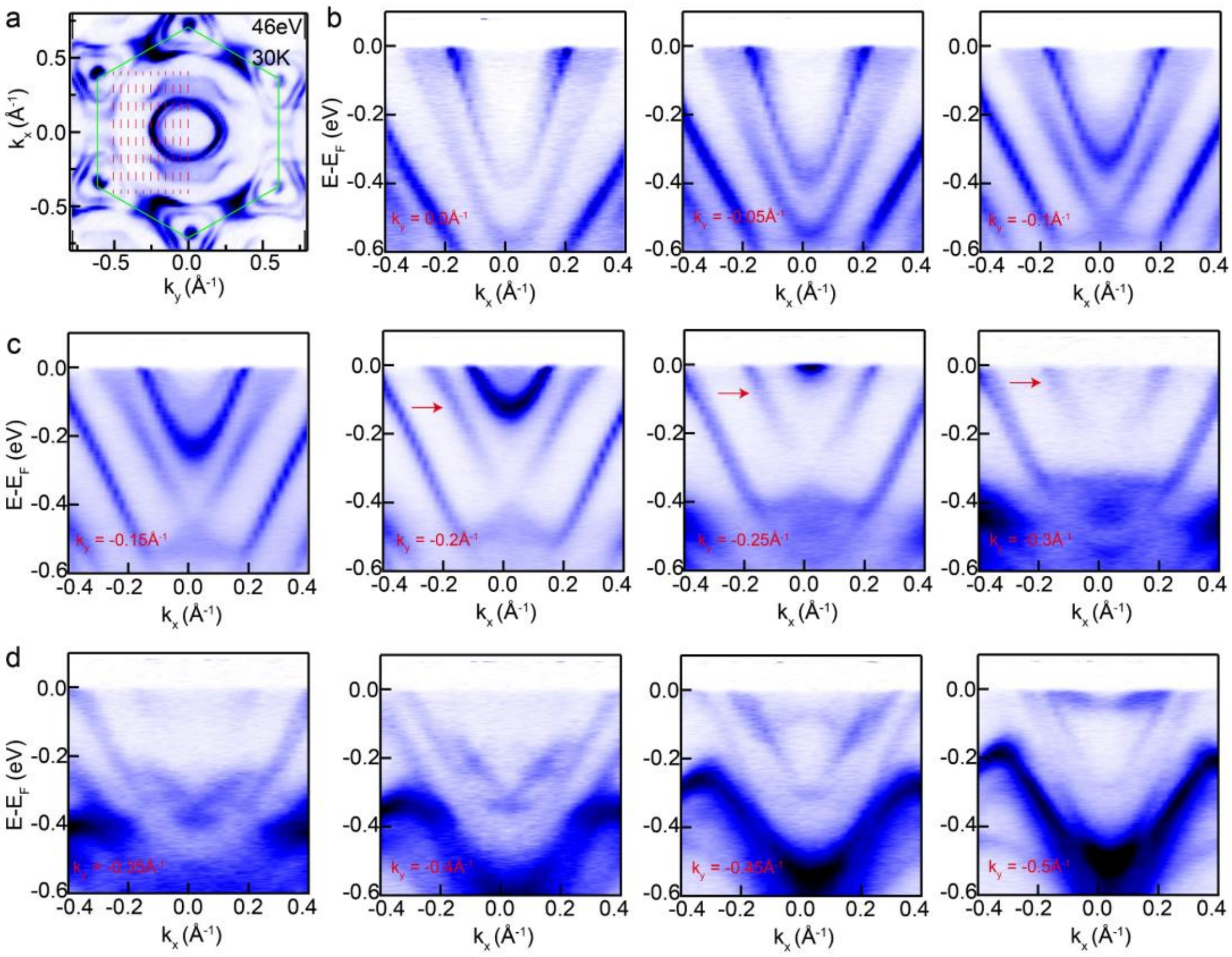

**Supplementary Figure 16 | The disappearing of the hybridization gap in the paramagnetic state of $TbTi_3Bi_4$ measured with photon energy 46 eV, LH polarization, at 30 K.** (**a**) The Fermi surface map of $TbTi_3Bi_4$ measured with photon energy 46 eV, LH polarization, at 6.3 K. (**b–d**) The ARPES intensity plots along the red dashed line in (**a**), from right to left. We can see that the hybridization gap is absent at all cuts.

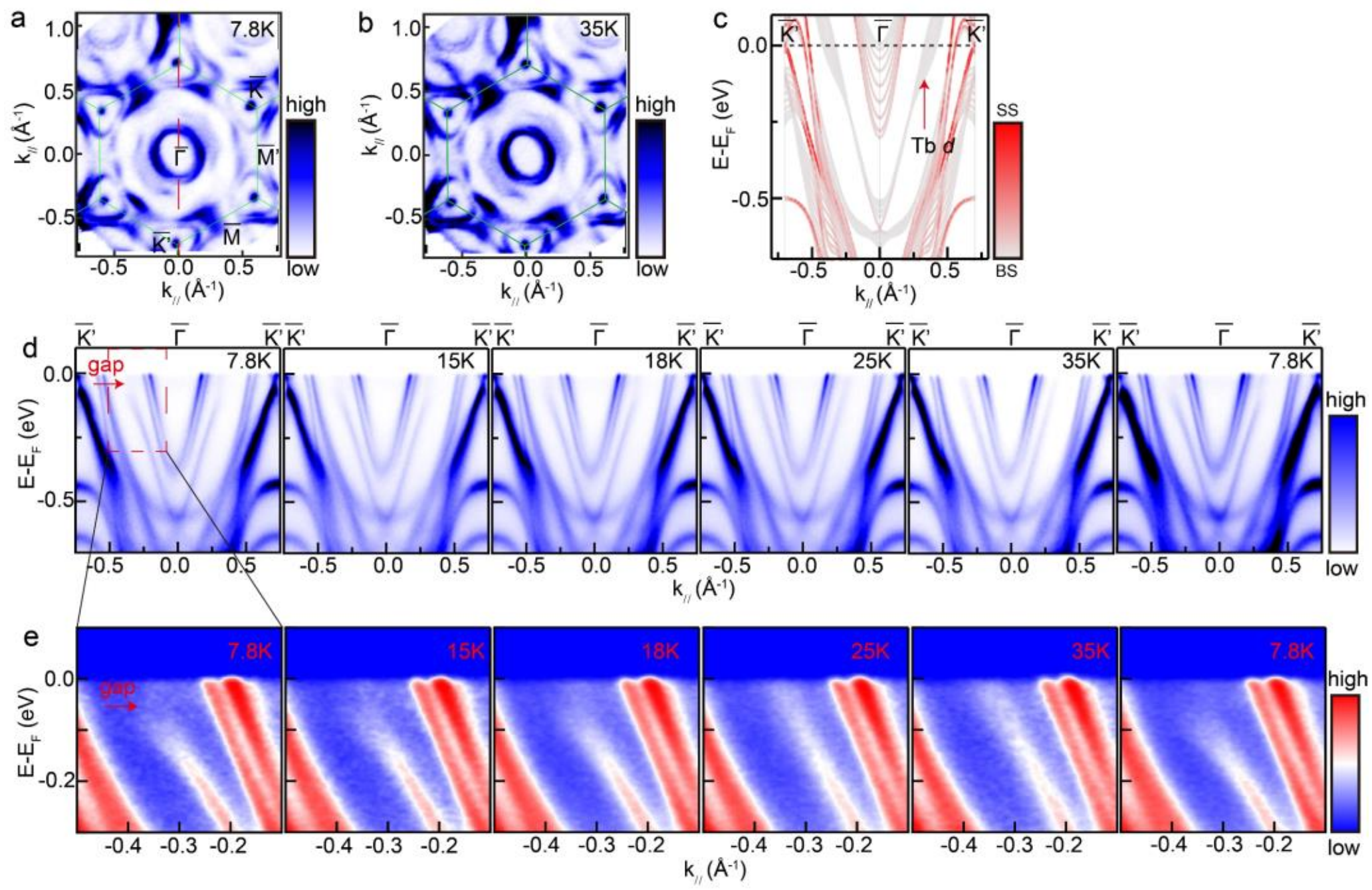

**Supplementary Figure 17 | The hybridization gap along the $\overline{K'}-\overline{\Gamma}-\overline{K'}$ direction.** (**a**, **b**) The

Fermi surfaces of $TbTi_3Bi_4$ measured with photons energy 46 eV, with LH polarization, at 7.8 K and 35 K respectively. (**c**) The calculated band projection along the $\overline{\mathrm{K}}' - \overline{\Gamma} - \overline{\mathrm{K}}'$ direction in the paramagnetic state. (**d**) The intensity plot of the measured dispersion along the $\overline{\mathrm{K}}' - \overline{\Gamma} - \overline{\mathrm{K}}'$ direction measured with photon energy 46 eV, with LH polarization, at various temperatures. (**e**) The zoom-in of the red dashed rectangular box in (**d**), showing the existence of the band gap extending to above $E_F$. Such bandgap closes in the paramagnetic state and reappears when cooling down to AFM state once again.

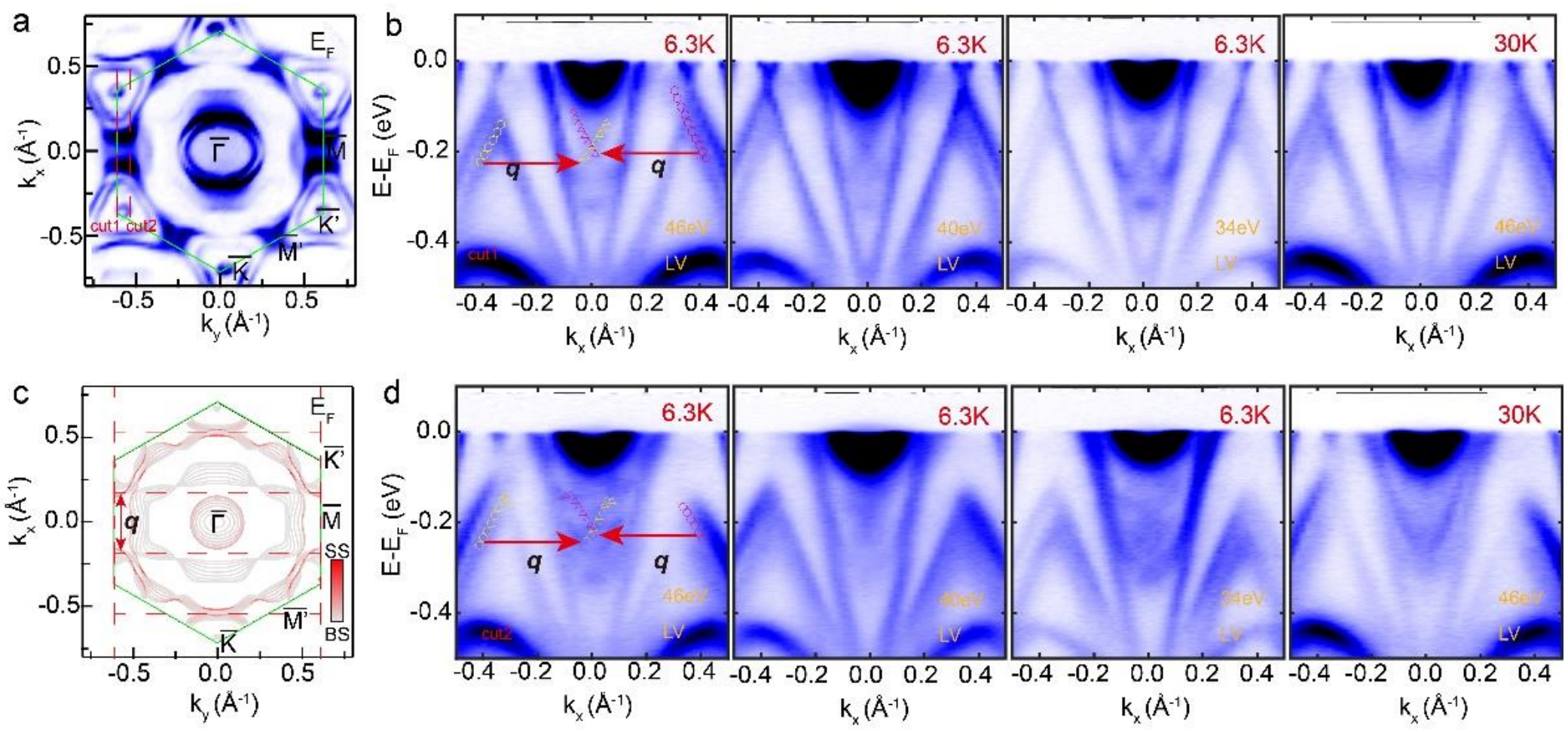

**Supplementary Figure 18 | The folded band along the $\overline{\mathrm{K}'} - \overline{\mathrm{M}} - \overline{\mathrm{K}'}$ direction measured with LV polarization. (a)** The Fermi surface of $TbTi_3Bi_4$ measured with photons energy 46 eV, with LV polarization at 6.3 K. Cut 1, 2 label the cut along and near the $\overline{\mathrm{K}}' - \overline{\mathrm{M}} - \overline{\mathrm{K}}'$ direction shown in **(b), (d)**. **(b)** The intensity plot of the measured dispersion along the cut 1 direction labelled in (a), the cuts are measured with various photon energies, with LV polarization, at different temperatures. Clear band folding of the Dirac band is shown at the $\overline{\mathrm{M}}$ point at 6.3 K. Triangles and circles label the main and folded bands connected by the folding vector **q**. At 30K, the band folding disappears. **(c)** Schematic diagram illustrating the folding vector **q** and the cut along the $\overline{\mathrm{K}}' - \overline{\mathrm{M}} - \overline{\mathrm{K}}'$ direction. **(d)** Same as (b), but along the cut 2 direction labelled in (a).

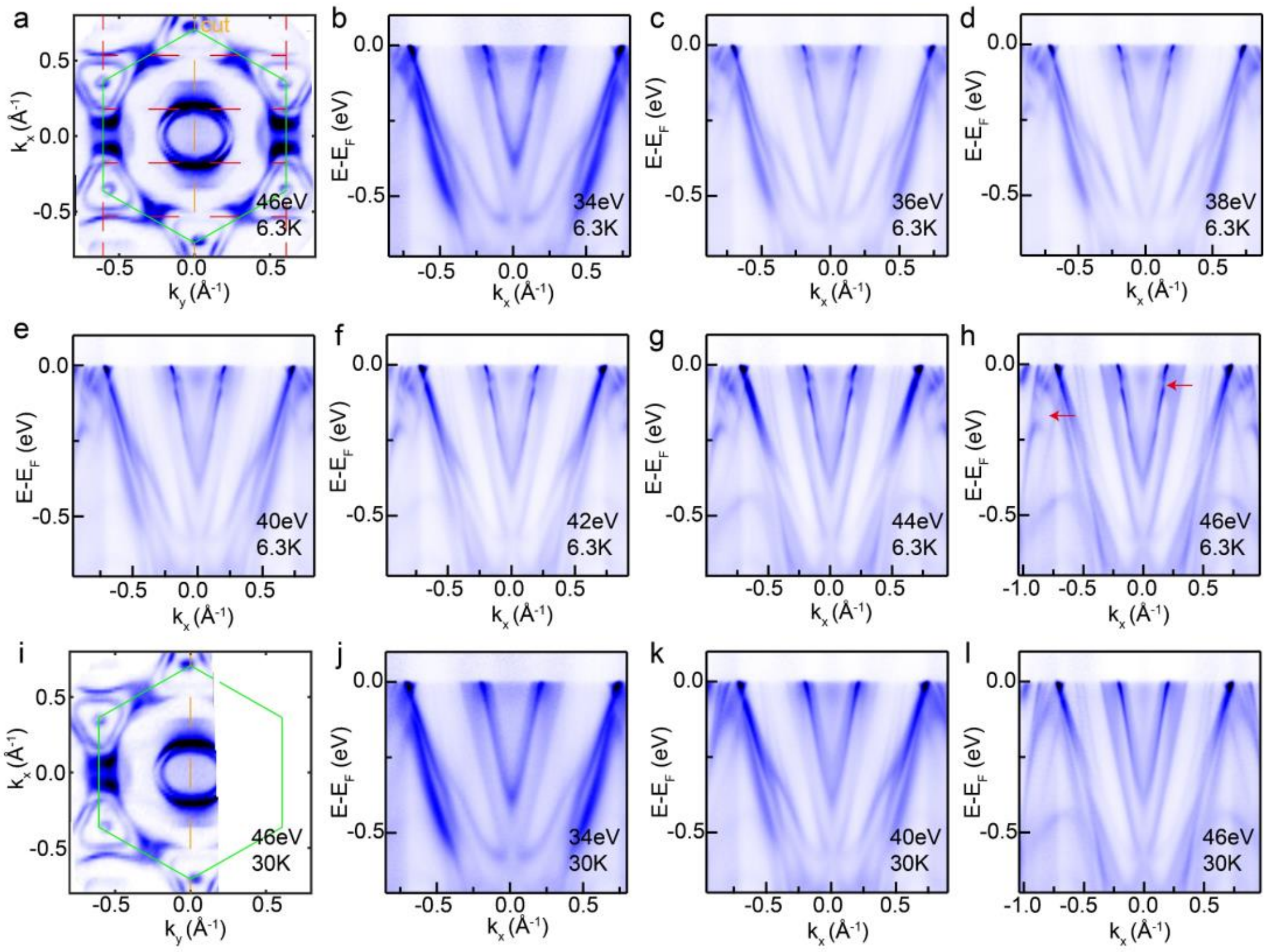

**Supplementary Figure 20 | The hybridization gap evolution near $\overline{\Gamma}$ along the $\overline{\mathrm{K}}-\overline{\Gamma}-\overline{\mathrm{K}}$ direction measured with LV polarization.** (**a**, **i**) The Fermi surface of $TbTi_3Bi_4$ measured with photon energy 46 eV at 6.3 and 30 K, respectively. (**b**–**h**) The ARPES intensity plots along the orange dashed line in (**a**) measured with different photon energies, corresponding to different $k_z$ values. (**j-l**) The ARPES intensity plots along the orange dashed line in (**i**) measured with photon energy 34, 40 and 46 eV, at 30 K. According to the $k_z$ measurement in Supplementary Fig. 9, the photon energy 34 eV corresponds to the $k_z = \pi$ plane and 46 eV corresponds to $k_z = 0$ plane. Therefore, we can see that the hybridization gap marked by red arrow in (**h**) exist at all $k_z$. These hybridization gap disappears in the paramagnetic state.

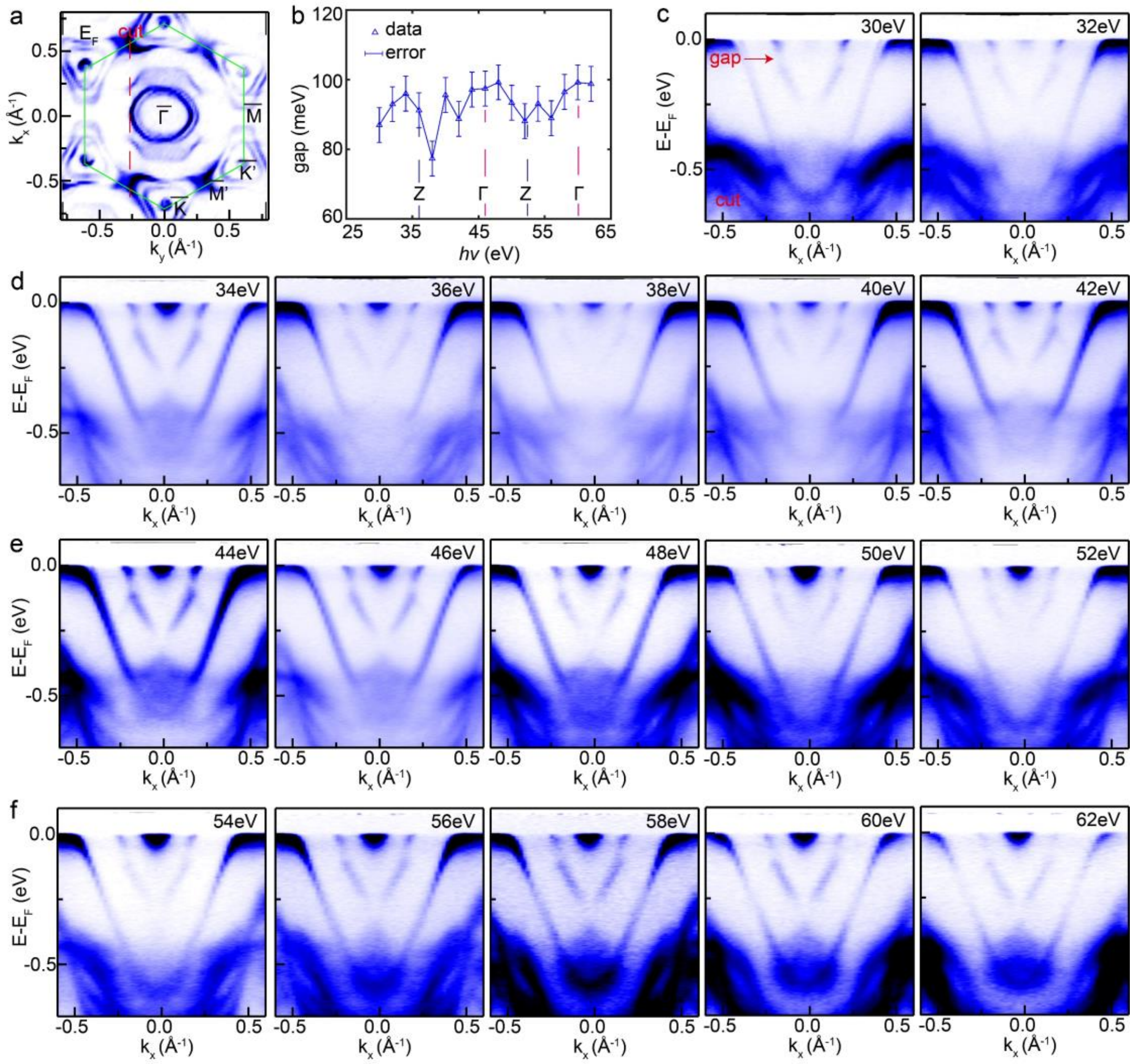

**Supplementary Figure 21| The hybridization gap evolution along $k_z$ direction measured at 6.3 K with LH polarization.** (**a**) The Fermi surface of $TbTi_3Bi_4$ measured with photon energy 46 eV and LH polarization. (**b**) The extracted gap evolution along $k_z$ direction, which shows periodic changes along $k_z$ direction. (**c–f**) The ARPES intensity plots along the red dashed line in (**a**) measured with different incident photons. Hybridization gap is indicated by the red arrow.

## Supplementary Note 8: The hybridization gap evolution with temperature

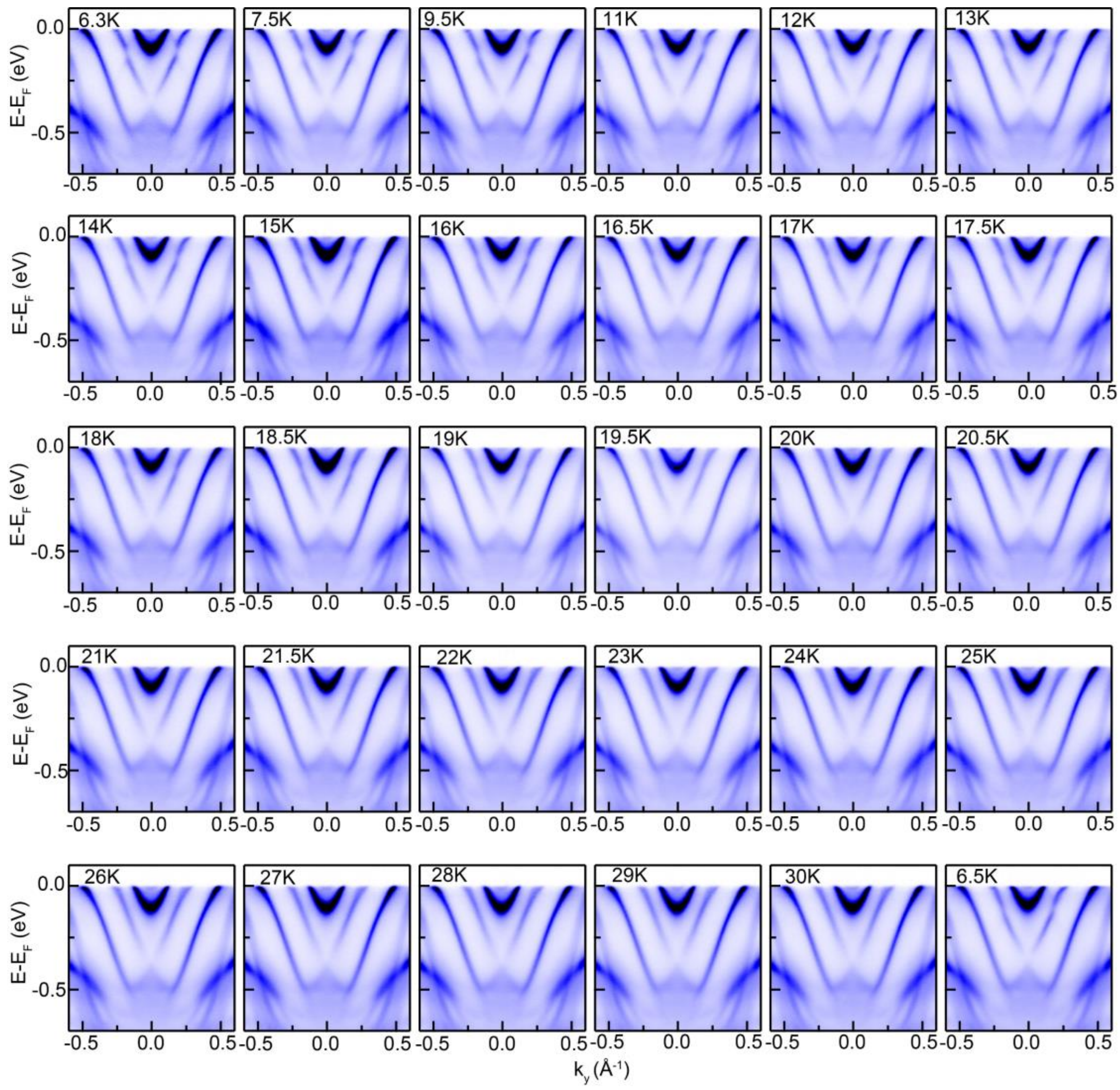

**Supplementary Figure 22 | The hybridization gap evolution with temperature along the black line in Figure 3c(i) of the main text, measured with LH polarization.** The measurement temperature was raised from 6.3K to 30 K and cooled back to 6.5 K.

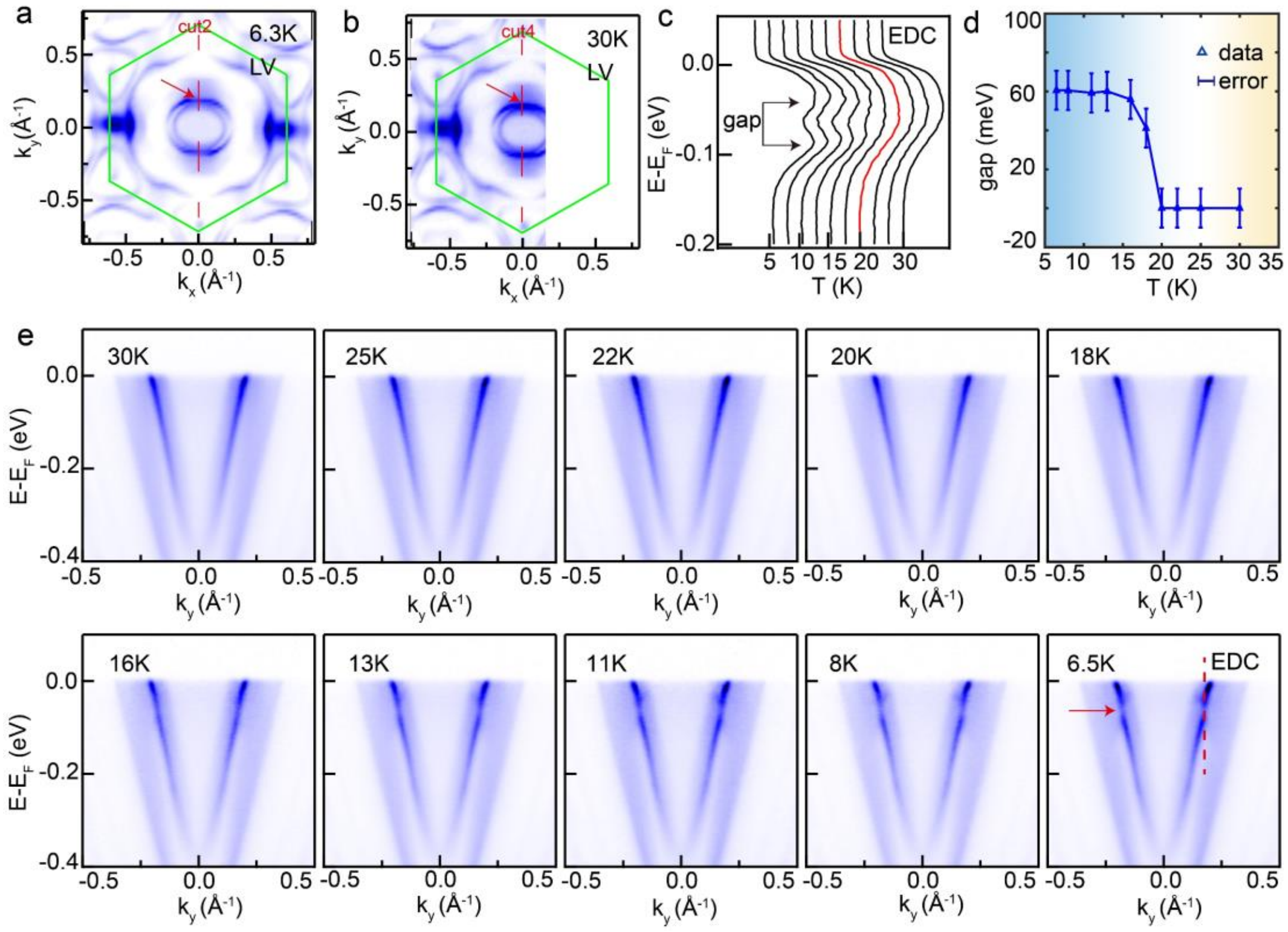

**Supplementary Figure 23 | The hybridization gap evolution near $\overline{\Gamma}$ with temperature, measured with 46 eV photons, LV polarization.** (**a**, **b**) The constant energy plane near $E_F$ measured at 6.3 K and 30 K, respectively. (**c**) The energy distribution curves (EDCs) along the red dashed line in (**e**) at different temperatures. Showing the gap deceases as temperature increases and merges near 20 K (the red curve). (**d**) The extracted gap size evolution in (**c**) showing the gap is about 60 meV at 6.5 K and closes near 20 K. (**e**) The ARPES intensity plots along the red dashed line in (**a**) measure with different temperature. The hybridization gap is marked with red arrow.

## Supplementary Note 9: The electronic structure of $TbTi_3Bi_4$ measured below $T_{N2}$

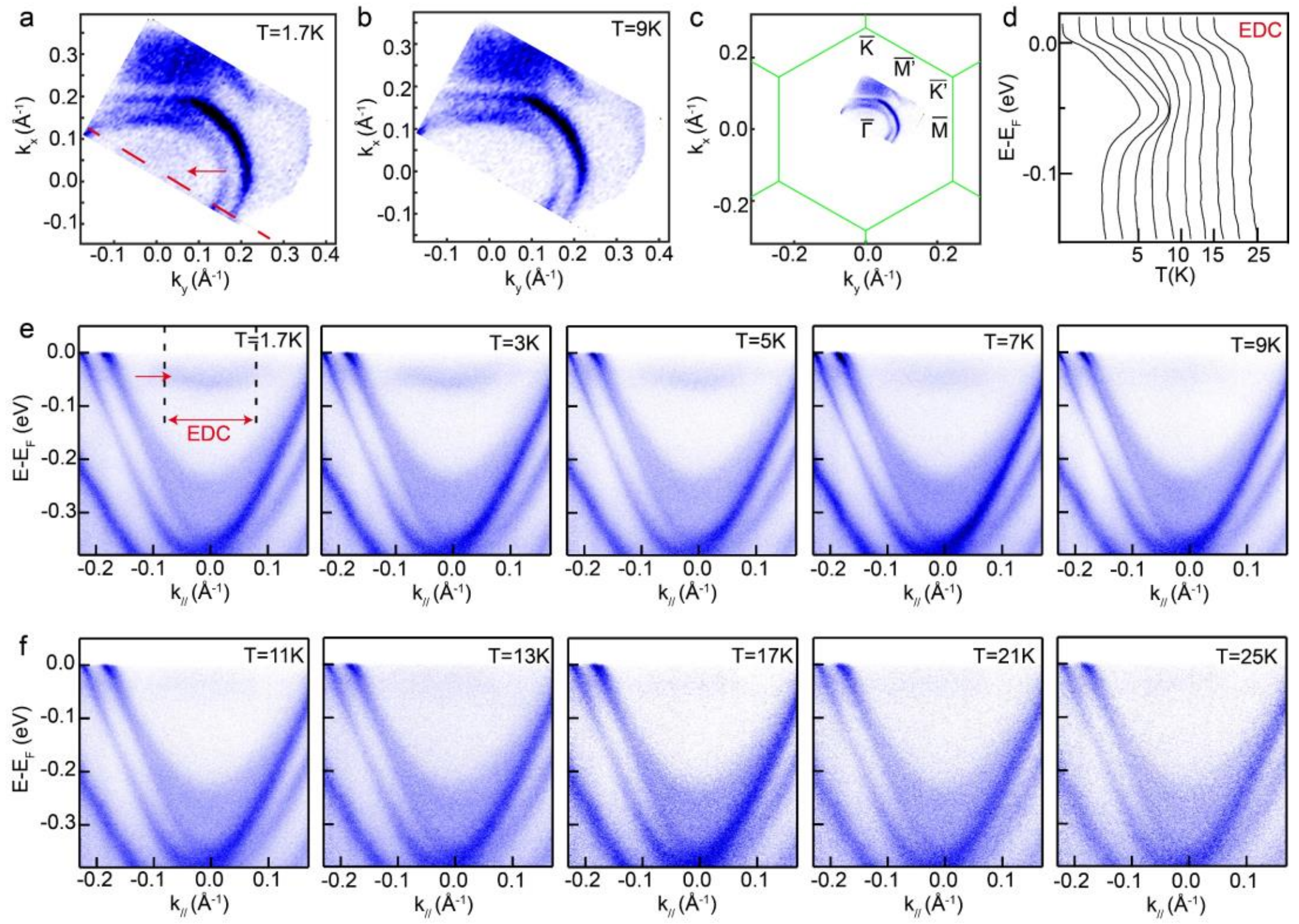

**Supplementary Figure 24 | Observation of a new band near Γ below $T_{N2}$. (a, b)** The Fermi surface of $TbTi_3Bi_4$ measured with 7 eV laser at 1.7 and 9 K respectively. **(c)** The illustration of the reciprocal space probed with 7 eV laser source. **(d)** The integrated EDC near Γ (as shown by the momentum window in (**e**)) with different temperatures. (**e, f**) The ARPES intensity plots along the red dashed lines in (**a**) at different temperatures. A new band is found near Γ (marked by red arrow in **e**) with temperature below $T_{N2}$ and gradually disappears around 10 K. This band can be seen in the Fermi surface in (**a**) as indicated by the red arrow.

## Supplementary Note 10: Additional scanning tunnelling microscopy data on $TbTi_3Bi_4$

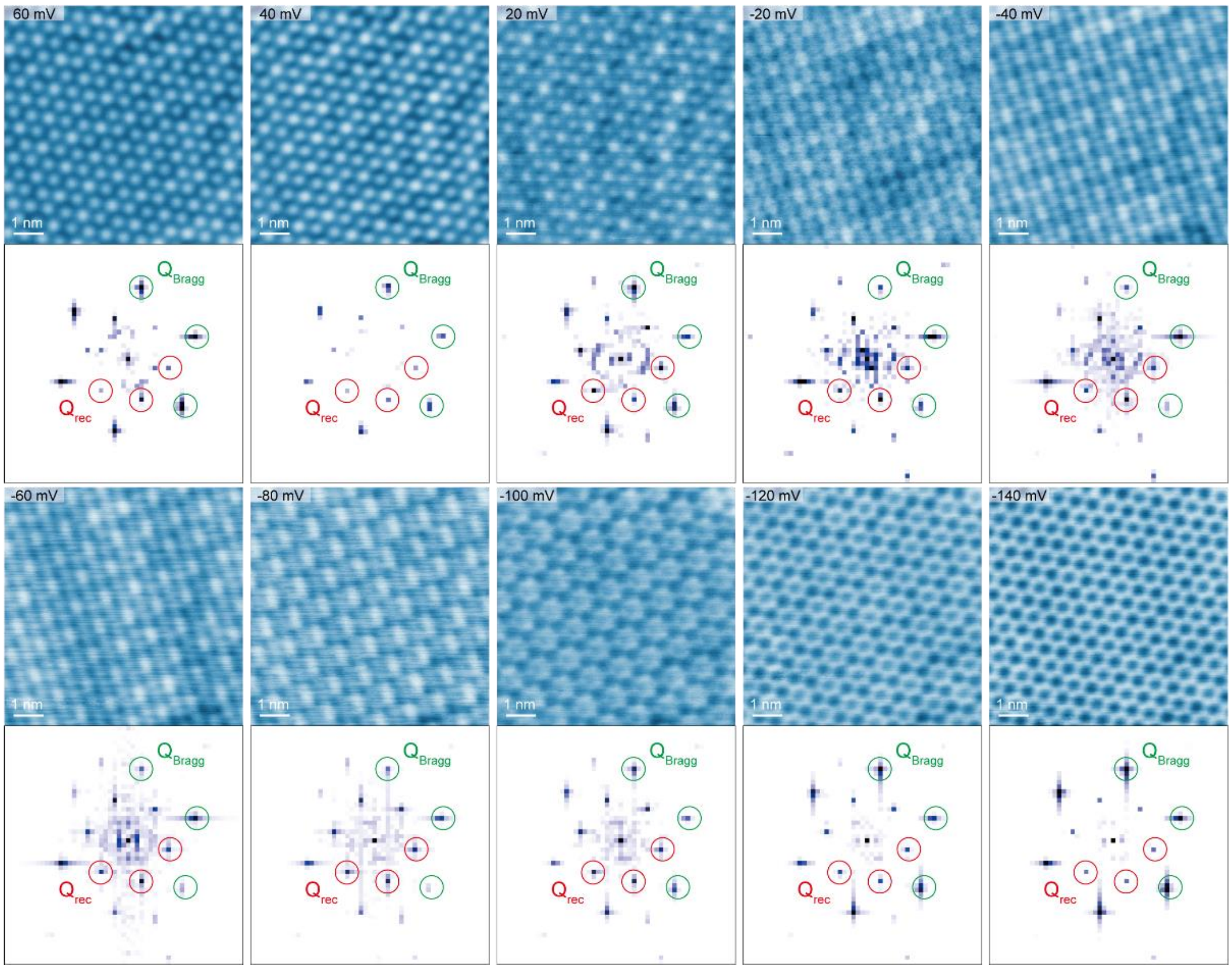

**Supplementary Figure 25 | The topographic images and corresponding Fourier transforms of $TbTi_3Bi_4$ with different bias voltages.** Atomically resolved topographic images of $TbTi_3Bi_4$ with different bias voltages. Scale bar, 1 nm. The green circles denote the primitive lattice, while the red circles indicate the in-plane additional ordering vector ($Q_{rec}$) of 3×1.

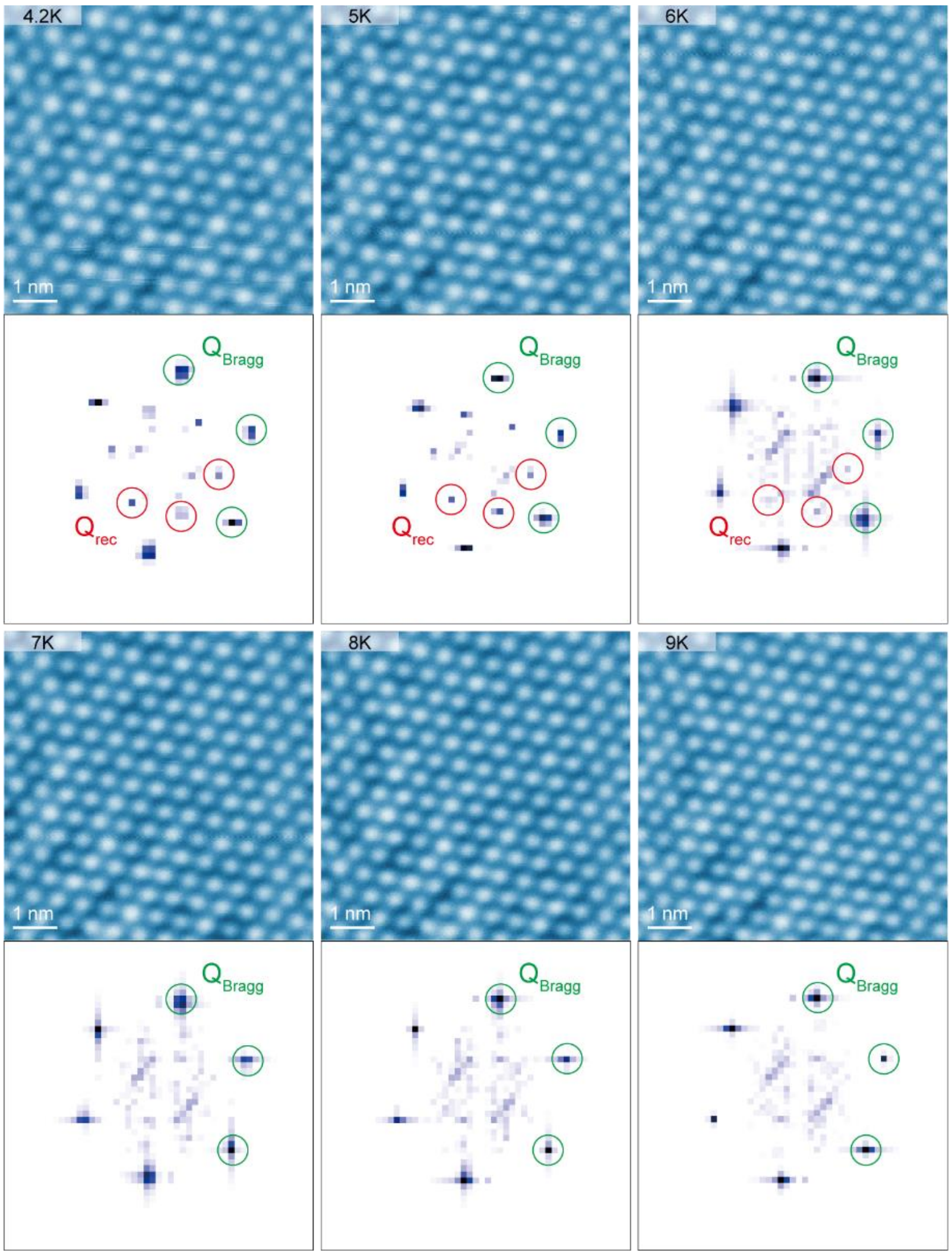

**Supplementary Figure 26 | Temperature-dependent topographic images and their corresponding Fourier transforms of $TbTi_3Bi_4$.** Atomically resolved topographic images of $TbTi_3Bi_4$ at different temperatures. Scale bar, 1 nm. The green circles denote the primitive lattice, while the red circles indicate the in-plane additional ordering vector ($Q_{rec}$) of 3×1.

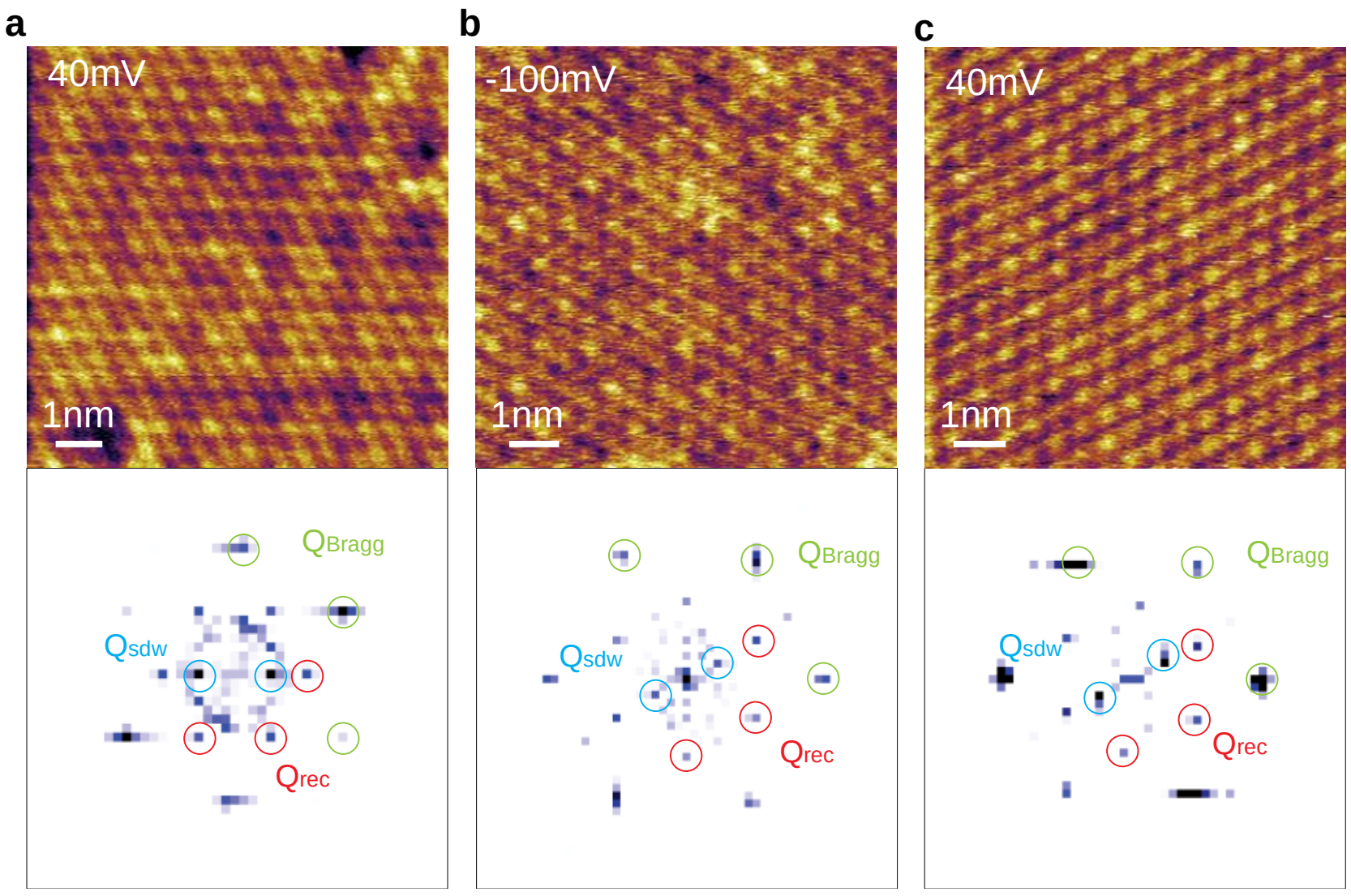

**Supplementary Figure 27 | Spin-resolved STM with different bias voltages at 4.55 K**. The *dI*/*dV* images of $TbTi_3Bi_4$ at different locations and their corresponding Fourier transforms. The green circles highlight the Bragg peaks of surface Tb lattice, the red circles highlight the in-plane structural reconstruction, and the blue circles highlight the SDW peaks of magnetic modulations. (**a**) *dI*/*dV* image obtained at the same location of 9 × 9 nm² area as Fig. 3 in main text. The scan window is rotated 30° from that of Fig. 3. Tunneling condition, Bias = 40 mV, $I_{set}$ = 1 nA. (**b,c**) *dI*/*dV* images obtained at different locations of 9 × 9 nm² area and their corresponding Fourier transform. Tunneling condition: (**b**) Bias = −100 mV, $I_{set}$ = 1 nA, (**c**) Bias = 40 mV, $I_{set}$ = 100 pA.

## Supplementary Note 11: Low-temperature transmission electron microscopy

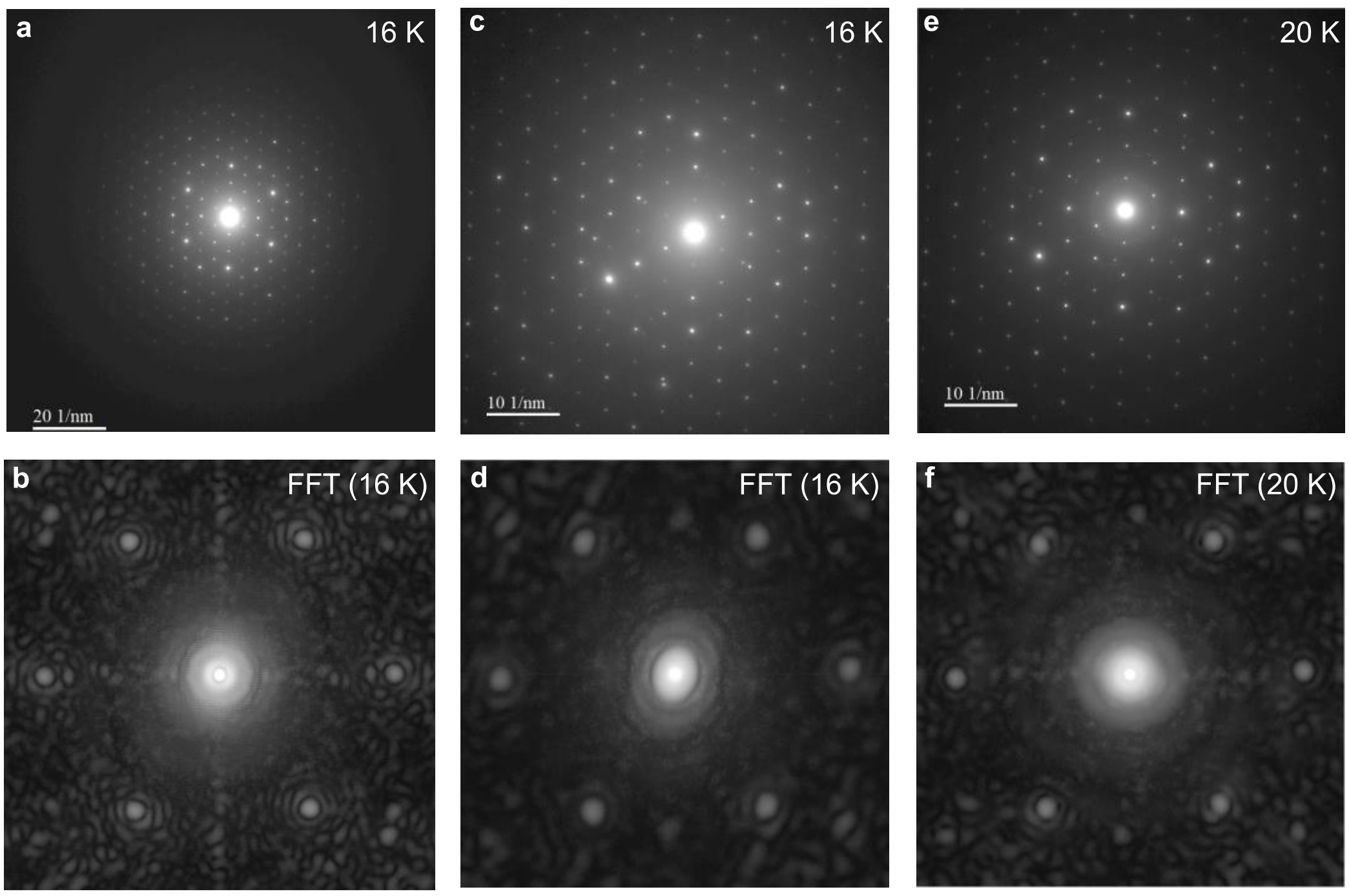

**Supplementary Figure 28 | Low-temperature transmission electron microscopy (TEM). (a,c)** TEM images of $TbTi_3Bi_4$ single crystal measured at 16 K with different scale bars. **(e)** TEM image at 20 K. (**b, d, f**) FFT patterns for (**a, c, e**), respectively. Down to 16 K, no bulk reconstruction can be distinguished.

## Supplementary Note 12: Thermal expansion and magnetostriction measurements of $TbTi_3Bi_4$

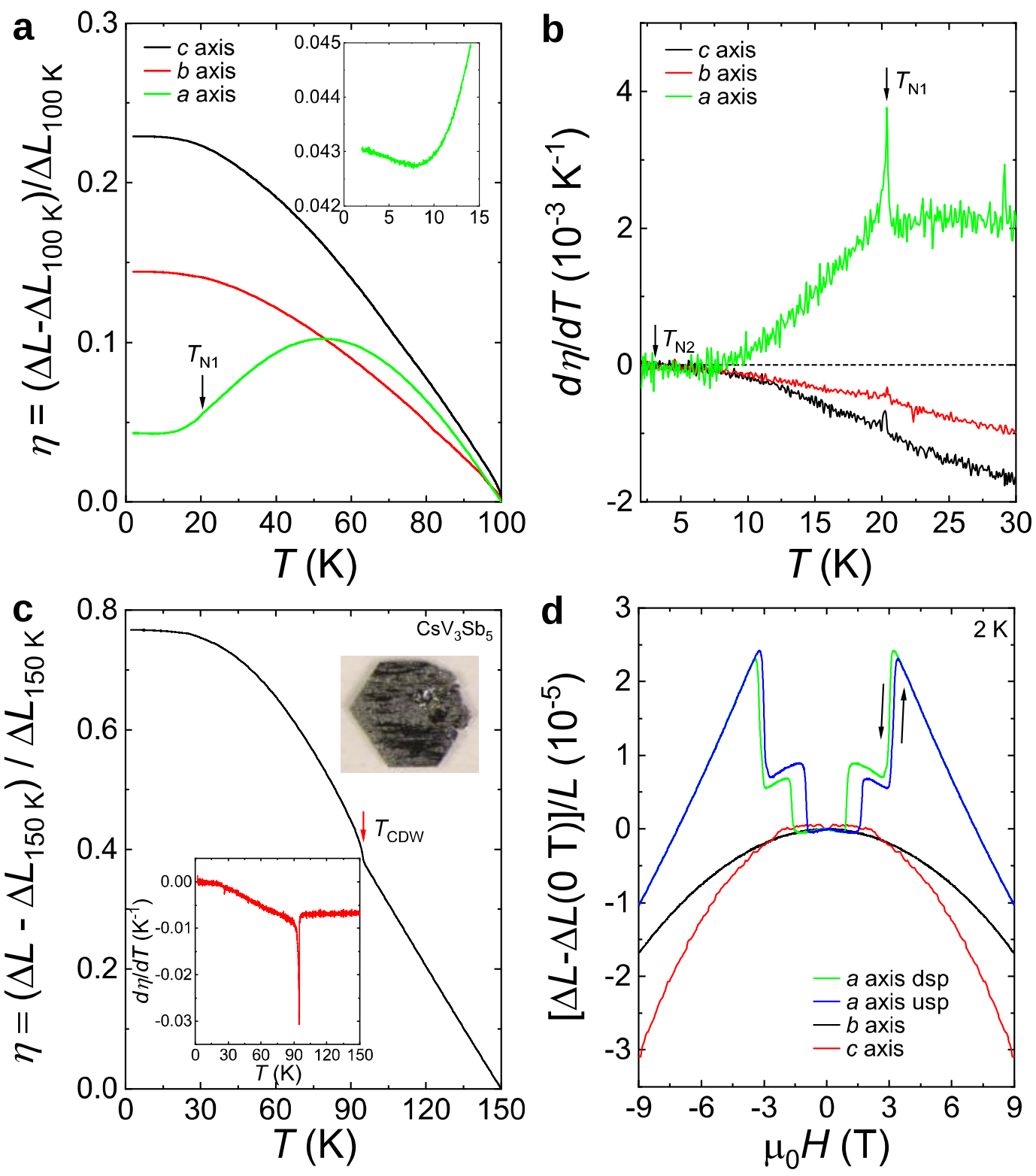

**Supplementary Figure 29 | Thermal expansion and magnetostriction experiments on$TbTi_3Bi_4$.** (**a**) Normalized relative length change, defined as $\eta = (\Delta L - \Delta L_{100\ K})/\Delta L_{100\ K}$, of a single crystal clamped along different orientations. (**b**) The normalized thermal expansion coefficient $d\eta/dT$ at low temperature. Black arrow represents the magnetic transition at $T_{N1}$. (**c**) Normalized relative length change of a single crystal of $CsV_3Sb_5$ clamped along the *c* axis for comparison. The red arrow represents the CDW transition. The lower inset shows the normalized thermal expansion coefficient $d\eta/dT$ at low temperature. The upper inset displays the measured sample. (**d**) Relative length change with respect to sample length, $[\Delta L - \Delta L(0\ T)]/L$, of a single crystal of $TbTi_3Bi_4$ clamped along different axis. Magnetic field is applied along the clamped direction.

Supplementary Fig. 29a presents the normalized relative length change profiles, defined as $\eta = (\Delta L - \Delta L_{100\ K})/\Delta L_{100\ K}$, of a $TbTi_3Bi_4$ single crystal clamped along different orientations. For the *b* and $c$ axis, $\eta$ monotonously increases with decreasing temperature, and saturate below ~ 20 K. However, $\eta$ displays nonmonotonic behavior for the $a$ axis. It peaks at ~ 55 K, and then shows a minimum at 8 K (the inset). At the magnetic transition temperature ~ $T_{N1}$, all axes exhibit weak kinks, and these kinks are more prominent in the normalized thermal expansion coefficient $d\eta/dT$ (Supplementary Fig. 29b). At $T_{N2}$, the anomaly in $\eta$ is very weak. As a comparison, a typical $CsV_3Sb_5$ single crystal has been measured (Supplementary Fig. 29c). The overall normalized relative length change of $CsV_3Sb_5$ displays a similar behavior (monotonic increase with decreasing temperature, and low temperature saturation). Especially at ~ 95 K where a CDW transition takes place, a distinct kink can be identified. The similar kink at $T_{N1}$ identified in $TbTi_3Bi_4$ suggests the bulk charge ordering occurs at the same temperature of the magnetic phase transition.

The relative length change with respect to sample length along different axis in magnetic fields has also been explored, as shown in Supplementary Fig. 29d. For both $b$ and $c$ axis, the relative length change is negative compared with that in zero field, while it overall displays positive behavior below the saturated field with a noteworthy hysteresis for the $a$ axis, consistent with magnetization curves. Beyond 3 T, the relative length change linearly decreases with fields. The field dependence of the length change indicates that magnetism is strongly coupled with lattice in $TbTi_3Bi_4$, and this strong coupling may give rise to prominent electronic structure evolution or even induce the observed SDW.

## Supplementary References

[1] Yin, J.-X., Lian, B. & Hasan, M. Z. Topological kagome magnets and superconductors. *Nature* **612**, 647–657 (2022).

[2] Jiang, Y.-X. *et al*. Unconventional chiral charge order in kagome superconductor $KV_3Sb_5$. *Nat. Mater.* **20**, 1353–1357 (2021).

[3] Wang, Z. *et al*. Electronic nature of chiral charge order in the kagome superconductor $CsV_3Sb_5$. *Phys. Rev. B* **104**, 075148 (2021).

[4] Jiang, K. *et al*. Kagome superconductors $AV_3Sb_5$ (A = K, Rb, Cs). *Natl. Sci. Rev.* **10**, nwac199 (2023).

[5] Korshunov, A. *et al*. Softening of a flat phonon mode in the kagome $ScV_6Sn_6$. *Nat. Commun.* **14**, 6646 (2023).

[6] Teng, X. *et al*. Discovery of charge density wave in a kagome lattice antiferromagnet. *Nature* **609**, 490–495 (2022).

[7] Teng, X. *et al*., Magnetism and charge density wave order in kagome FeGe. *Nat. Phys.* **19**, 814–822 (2023).

[8] Huang, X. *et al*. Observation of the chiral-anomaly-induced negative magnetoresistance in 3D

Weyl semimetal TaAs. *Phys. Rev. X* **5**, 031023 (2015).

[9] Kuroda, K. *et al*. Evidence for magnetic Weyl fermions in a correlated metal. *Nat. Mater.* **16**, 1090–1095 (2017).

[10] Cheng, E. *et al*. Magnetism-induced topological transition in $EuAs_3$. *Nat. Commun.* **12**, 6970 (2021).